\documentclass[journal]{IEEEtran}  
\ifCLASSINFOpdf
\else
\fi

\usepackage{subcaption}
\usepackage{color}
\usepackage[pdfpagelabels,breaklinks,
pdfauthor={Korobkov A. Alexey, Marina~K.~Diugurova, Jens~Haueisen, Martin~Haardt},
pdftitle={Robust Multi-dimensional Model Order Estimation Using LineAr Regression of Global Eigenvalues (LaRGE)},
pdfkeywords={eigenvalue, global eigenvalue, tensor, the rank of the tensor, the model order of multi-dimensional data}]{hyperref} 
\usepackage{comment}
\usepackage{mathtools}
\usepackage{float}

\newcommand{\bs}{\boldsymbol}
\newcommand{\ts}[1]{\bs{\mathcal{#1}}}

\usepackage{epsfig}
\usepackage{bmpsize}

\usepackage[yyyymmdd,hhmmss]{datetime}

\usepackage{amsmath,amssymb,amsfonts}   
\usepackage{graphicx}                   
\usepackage{epstopdf}
\usepackage{psfrag}                     
\usepackage{tikz}
\usepackage{pgfplots}
\usepackage{amsmath,graphicx}
\usepackage{multirow}
\usepackage{color}
\usepackage{bm,MnSymbol}

\newcommand{\multilinecomment}[1]{}

\renewcommand{\rm}{\textrm}

\newcommand{\ma}[1]{\bm{{#1}}}         
\newcommand{\ten}[1]{\ma{{\mathcal{#1}}}} 

\newcommand{\compl}{\mathbb{C}}        
\newcommand{\real}{\mathbb{R}}         

\newcommand{\opof}[2]{\mathop{{\rm{#1}}}\left(#2\right)}         

\newcommand{\minoff}[1]{\opof{min}{#1}}           

\newcommand{\normof}[2]{\left\|#1\right\|_{#2}}
\newcommand{\fronorm}[1]{\normof{#1}{\rm F}}            
\newcommand{\honorm}[1]{\normof{#1}{\rm H}}               

\newcommand{\trans}{{\rm T}}   
\newcommand{\herm}{{\rm H}}    

\newcommand{\unfnot}[2]{\left[ #1 \right]_{(#2)}}    
\newcommand{\unf}[2]{\unfnot{\ten{#1}}{#2}}          

\usepackage{color}

\begin{document}
	
	\pagenumbering{gobble}
%


\title{Robust Multi-dimensional Model Order Estimation Using LineAr Regression of Global Eigenvalues (LaRGE)}


%
%
%

\author{Alexey~A.~Korobkov,
        Marina~K.~Diugurova,
        Jens~Haueisen,~\IEEEmembership{Member,~IEEE},
        and Martin~Haardt,~\IEEEmembership{Fellow,~IEEE}
\thanks{Alexey~A.~Korobkov and Marina~K.~Diugurova are with the Department for Radio-Electronic and Telecommunication Systems, Institute for Radio-Electronics, Photonics, and Digital Technologies, Kazan National Research Technical University n.a. A.N Tupolev-KAI, Kazan, Russia, e-mail: korobkov@inbox.ru.}
\thanks{Jens~Haueisen is with the Institute of Biomedical Engineering and Informatics, Ilmenau University of Technology, Ilmenau, Germany, email: jens.haueisen@tu-ilmenau.de.}
\thanks{Martin~Haardt is with the Communications Research Laboratory, Ilmenau University of Technology, Ilmenau, Germany, email: martin.haardt@tu-ilmenau.de.}}
\maketitle

\begin{abstract}
The efficient estimation of an approximate model order is very important for real applications with multi-dimensional data if the observed low-rank data is corrupted by additive noise. In this paper, we present 
a novel robust method for model order estimation of noise-corrupted multi-dimensional low-rank data based on the LineAr Regression of Global Eigenvalues (LaRGE). The LaRGE method uses the multi-linear singular values obtained from the HOSVD of the measurement tensor to construct global eigenvalues. In contrast to the Modified Exponential Test (EFT) that also exploits the approximate exponential profile of the noise eigenvalues, LaRGE does not require the calculation of the probability of false alarm. Moreover, LaRGE achieves a significantly improved performance in comparison with popular state-of-the-art methods. It is well suited for the analysis of biomedical data. The excellent performance of the LaRGE method is illustrated via simulations and results obtained from EEG recordings.
\end{abstract}


%
\IEEEpeerreviewmaketitle

%
%
%
%

\section{Introduction}

\IEEEPARstart{M}{ulti-dimensional models} are widespread in a variety of applications, for example, radar, sonar, channel modeling in wireless communications, image processing, the estimation of MIMO channels parameters, blind source separation and many more \cite{7038247}. According to these models the  measured signals or the data can be stacked into multi-dimensional arrays or tensors. Moreover, in biomedical data processing multi-dimensional models have been widely used recently. For example, biomedical signals like Electroencephalograms (EEG), Magnetoencephalograms (MEG) or Electrocardiograms (ECG) are recorded from many sensors simultaneously. Therefore, it is natural to use multi-dimensional models or tensors for representing these signals. 

Different types of tensor decompositions are used for the extraction of features from the data or to denoise recorded signals. However, in biomedical signal processing the most frequently used decompositions of multi-dimensional data are an approximate low-rank Canonical Polyadic (CP) decomposition \cite{KoBa09} also known as Parallel Factor (PARAFAC) analysis \cite{Har70} or Canonical Decomposition (CANDECOMP) \cite{CarrCh70} and the truncated Multi-Linear Singular Value Decomposition (MLSVD) \cite{Lathauwer2000AMS} also known as Higher Order Singular Value Decomposition (HOSVD).

According to the CP model, a tensor is decomposed into the minimum number $R$ of rank-one components. Hence, the proper choice of the model order affects the accuracy of the processing and subsequently also the interpretable results. This problem is very important when measured biomedical data are processed. However, in most cases, the observed data are corrupted by noise. Therefore, the problem of estimating the order of an approximate low-rank model is a non-trivial task.

The first attempts to develop methods for overcoming this problem were made at the beginning of the 1970s. In \cite{AKAIKE} Akaike's information criterion (AIC) was proposed. This criterion takes the observed data structure into account. In \cite{Rissanen1978} another criterion was proposed that penalizes the over parameterization more strongly than AIC. Schwarz proposed a Bayesian information criterion (BIC) for the estimation of the order of linear models that describe independent and identically distributed observations \cite{Schwarz1978}. Wax and Kailath implemented the AIC and the Minimum Description Length (MDL) scheme that has been derived from the BIC for the detection of the number of signals in a multi-channel time series \cite{WaxTail1985}. However, these methods often fail when the number of the snapshots of the observed data is small. 

To overcome this problem, the Exponential Fitting Test (EFT) based on the geometrical profile of noise-only eigenvalues has been developed and presented in \cite{Quinlan2006ModelOS}. The EFT allows the probability of false alarm to be controlled and predefined, which is a crucial point for systems such as RADARs. Moreover, in \cite{CostaHaardtRoemer2007} the EFT was improved for multidimensional signals via the Modified Exponential Fitting Test (M-EFT) that outperforms all other schemes for the cases when the data are corrupted by Gaussian noise. Furthermore, the authors showed how the classical AIC and MDL methods can be improved. In \cite{Joao2011} the authors presented the $N$-dimensional ($N$-D) EFT that outperforms the other techniques for cases with Gaussian noise. For colored noise a new Closed-Form PARAFAC-based Model Order Selection (CFP-MOS) was proposed. Moreover, the AIC and MDL techniques were extended to the multi-dimensional cases. Several methods for estimating the order of CP models were presented in \cite{Liu2016DetectionON}, \cite{7676397}, \cite{Pouryazdian2016CANDECOMPPARAFACMO}. In \cite{SalvadorChan2004}, the authors propose a technique to find the "knee" of an error curve to determine the number of clusters. This technique can also be used to estimate the model order. In \cite{Yu2021}, the authors implemented the $k$-means clustering algorithm to separate the 1-mode singular values of the covariance tensor into two clusters: the signal cluster with large singular values and the noise cluster with the remaining smaller singular values. The number of elements in the signal cluster is assumed as the model order.

The estimation of the model order in biomedical signal processing problems is often based on the results of a visual inspection of the singular values or some assumptions about the structure of the data. To avoid this visual inspection and to provide a systematic approach for model order estimation in practical applications, we propose a novel robust method based on the LineAr Regression of Global Eigenvalues (LaRGE). In addition to the initial results presented in \cite{KorDiugHaueisHaardt2020}, in this paper, we provide a more detailed derivation of the equations of the LaRGE method. Moreover, we introduce a heuristic alternative by incorporating a penalty function (LaRGE-PF) that shows an improved performance in the case of relatively small tensors. We compare the performance of LaRGE and LaRGE-PF with the classical AIC and MDL schemes and their $N$-dimensional extensions. Furthermore, we provide detailed results of the model order estimation and the subsequent low-rank tensor decomposition of measured EEG data during Intermittent Photic Stimulation.


In Section~\ref{sec:Data_model} the multi-dimensional data model is presented. Section~\ref{sec:The_Method} describes the proposed LaRGE and LaRGE-PF schemes. As a benchmark, extensions of two classical methods AIC and MDL (or BIC) to the multi-dimensional cases are reviewed in Section~\ref{sec:Classical_methods}. Section~\ref{sec:Results} describes the simulation scenarios and compares the performance of the LaRGE, LaRGE-PF, AIC, MDL, $N$-D AIC, and $N$-D MDL. 
Inspired by the good performance of LaRGE and LaRGE-PF, we also implemented them for model order estimation of EEG recordings. The results are presented in Section ~\ref{sec:ResultsRealData}. In Section~\ref{sec:Conclusion}, we conclude the paper.



\section{Data model and notation}
\label{sec:Data_model}

In this paper, we use the following notation, $a$, \textbf{a}, \textbf{A}, and $\ten{A}$ are used to denote scalars, column vectors, matrices, and tensors, respectively. Moreover, \textbf{a}($i$) defines the element ($i$) of a vector \textbf{a}. The same applies to a matrix \textbf{A} ($i,j$) and a tensor $\ten{A}$ ($i,j,k$). The tensor $\ten{I}_{D,R}$ is $D$-dimensional super-diagonal tensor of size $R \times R \times \ldots \times R$, which is equal to one if all $D$ indices are equal and zero otherwise. The $d$-mode product between a $D$-way tensor of size $M_d$ along mode $d=1, 2, \ldots, D$ represented as $\ten{A} \in \compl^{M_1 \times M_2 \times \dots \times M_D}$ and a matrix $\ma{U} \in \compl^{J \times M_d}$ is written as $\ten{A} \times_d \ma{U}$. It is computed by multiplying all $d$-mode vectors of $\ten{A}$ with $\ma{U}$, whereas the $d$-mode vectors of $\ten{A}$ are obtained by varying the $d$-th index from $1$ to $M_D$ and keeping all other indices fixed. Aligning all $d$-mode vectors as the columns of a matrix yields the $d$-mode unfolding of $\ten{A}$ which is denoted by $\unf{A}{d} \in \compl^{M_d \times M_{d+1} \cdot \ldots \cdot M_D \cdot M_1 \cdot \ldots \cdot M_{d-1}}$. 

The CP decomposition of a $D$-way noiseless tensor $\ten{X}_0 \in \compl^{M_1 \times M_2 \times M_3 \times \dots \times M_D} $ is represented as
\begin{align}
\ten{X}_0=\ten{I}_{D,R} \times_1 \ma{F}_1 \times_2 \ma{F}_2 \times_3 \ma{F}_3 \times \dots \times_D \ma{F}_D,
\label{eq:X0}
\end{align}
where $\ma{F}_d \in \compl^{M_d \times R} (d=1,2,3, \ldots D)$ are the factor matrices and $R$ is the order of the CP model or the rank of the tensor $\ten{X}_0$.

In practice, the recorded data are corrupted by noise. The tensor that is constructed from observations can be defined as
\begin{align}
\ten{X}=\ten{X}_0+\ten{N},
\end{align}
where $\ten{N} \in \compl^{M_1 \times M_2 \times M_3 \times \dots \times M_D}$ is the additive noise tensor.
Therefore, \eqref{eq:X0} can be rewritten as
\begin{align}
\ten{X}=\ten{I}_{D,R} \times_1 \ma{F}_1 \times_2 \ma{F}_2 \times_3 \ma{F}_3 \times \dots \times_D \ma{F}_D+\ten{N}.
\label{eq:X}
\end{align}

Obviously, the rank of the tensor $\ten{X}$ is not equal to $R$. In general, it is bigger. Therefore, after a CP decomposition of the tensor with observations $\ten{X}$, we obtain the estimates $\hat{\ma{F}}_d \in \compl^{M_d \times R} (d=1,2,3, \ldots D)$ of factor matrices $\ma{F}_d \in \compl^{M_d \times R} (d=1,2,3, \ldots D)$
\begin{align}
\ten{X} \approx \ten{I}_{D,R} \times_1 \hat{\ma{F}}_1 \times_2 \hat{\ma{F}}_2 \times_3 \hat{\ma{F}}_3 \times \dots \times_D \hat{\ma{F}}_D.
\label{eq:Xhat}
\end{align}

On the other side, the HOSVD model of a tensor $\ten{X}$ is defined by
\begin{align}
\ten{X}=\ten{S} \times_1 \ma{U}_1 \times_2 \ma{U}_2 \times_3 \ma{U}_3 \times \dots \times_D \ma{U}_D,
\label{eq:X_HOSVD}
\end{align}
where $\ten{S} \in \compl^{M_1 \times M_2 \times M_3 \times \dots \times M_D}$ is the core tensor and $\ma{U}_r \in \compl^{M_d \times M_d}, (d=1,2,3, \ldots D)$ are the unitary factor matrices. 

As shown in \cite{Quinlan2006ModelOS}, \cite{CostaHaardtRoemer2007}, \cite{Joao2011} the $d$-mode singular values that are contained in the core tensor $\ten{S}$ play a major role in the problem of multi-dimensional model order estimation. 
The $d$-mode singular values can be computed via the Singular Value Decomposition (SVD) of the $d$-mode unfolding of the tensor $\ten{X}$ according to
\begin{align}
\unf{X}{d}=\ma{U}_d \cdot \ma{\Sigma}_d \cdot \ma{V}_d^\herm,
\label{eq:UnfX}
\end{align}
where $\ma{U}_d \in \compl^{M_d \times M_d}$, $\ma{V}_d \in \compl^{\tilde{M}_d \times \tilde{M}_d}$ are unitary matrices and $ \ma{\Sigma}_d \in \compl^{M_d \times \tilde{M}_d}$ is a diagonal matrix that has the $d$-mode singular values $\sigma_i^{(d)}$ on the main diagonal and $\tilde{M}_d=\frac{M}{M_d}$.

\section{LaRGE method for multi-dimensional model order estimation}
\label{sec:The_Method}
\subsection{Global Eigenvalues}

As is shown in \eqref{eq:X_HOSVD}, the $d$-mode singular values represent the internal structure of the data and should be exploited for model order estimation. However, tensors with real data may have different $d$-ranks in different modes. Moreover, when an approximate low-rank CP decomposition is computed, an estimate of the rank of the noiseless model that is equal for all $d$-mode unfoldings has to be found. Therefore, we can define \textit{global eigenvalues} as proposed in \cite{CostaHaardtRoemer2007} and \cite{Joao2011} as the product of all $d$-mode singular values with the same indices as follows
\begin{align}
\tilde{\lambda}_i^{[G]}={ \prod_{d=1}^{D} \left( \sigma_i^{(d)} \right)^2 }, i=1,2,3, \ldots,M^{[G]},
\label{eq:GlobLamda}
\end{align}
where ${M^{[G]}}=\minoff{M_d}, d=1,2,3, \ldots,D$. 

Assuming that the global eigenvalues include the information about the signals and the noise, the set of global eigenvalues can be divided into two subsets. One subset contains the global eigenvalues that represent the noise in the observations. These global eigenvalues are called \textit{noise global eigenvalues}.  The other subset contains the \textit{signal global eigenvalues}. The number of the signal global eigenvalues corresponds to the model order or the rank of the tensor $ {\ten X}_0$ as well as the number of sources or components in the noisy observations.

\subsection{LaRGE method}

In \cite{CostaHaardtRoemer2007} and \cite{Joao2011}, it was shown that the noise global eigenvalues have approximately an exponential profile and there is at least one noise global eigenvalue in the set of all global eigenvalues.  Starting from the smallest global eigenvalue, the noise global eigenvalues follow a straight line on a logarithmic scale (due to their approximate exponential distribution)
\begin{align}
\label{eq:GlEigen}
\lambda_i^{[G]}=\ln \left( \tilde{\lambda}_i^{[G]} \right) . 
\end{align}

Therefore, we want to determine a linear approximation $\hat{\lambda}_i^{[G]} $ of the actual noise global eigenvalues profile $\lambda_i^{[G]}$ on a logarithmic scale based on the least squares criterion
\begin{align}
\minoff{\sum_{i=M^{[G]}}^{M^{[G]}-k} \left( \hat{\lambda}_i^{[G]} - \lambda_i^{[G]} \right)^2  },
\end{align}
where $ \hat{\lambda}_i^{[G]}=a_1 \cdot i + a_2, \; i=M^{[G]},M^{[G]}-1,M^{[G]}-2, \ldots,M^{[G]}-k$. We sequentially approximate the profile of the noise global eigenvalues via a straight line starting from the smallest global eigenvalues with indices $i = M^{[G]}$ and $i = M^{[G]} - 1$ for $k=1$, where the $k$ is the step index. Using this linear approximation the value of the next largest noise eigenvalue can be predicted.

In each step $ k = 1, 2, ...., M^{[G]}-1$ the absolute prediction error $\Delta_{i}^{[G]}$ is calculated according to
\begin{align}
\Delta_{M^{[G]}-k}^{[G]} = \Delta_{i}^{[G]} = \lambda_{i}^{[G]} - \hat{\lambda}_{i}^{[G]}.
\end{align}
A small value of the absolute prediction error indicates that a noise global eigenvalue is found. On the contrary, a big prediction error can be indicative of a signal global eigenvalue. 

To normalize the scale of values of the prediction errors, we define the relative prediction error
\begin{align}
\delta_{M^{[G]}-k} = \frac{\Delta_{M^{[G]}-k}^{[G]}}{\left| \hat{\lambda}_{M^{[G]}-k}^{[G]} \right| }=\dfrac{\lambda_{M^{[G]}-k}^{[G]} - \hat{\lambda}_{M^{[G]}-k}^{[G]}}{\left| \hat{\lambda}_{M^{[G]}-k}^{[G]}\right| }.
\label{eg:RelDelta}
\end{align}

We consider the standard deviation of the approximation error $\sigma_{M^{[G]}-k}^2$ to automatically find the first signal global eigenvalue
\begin{align}
\sigma_{M^{[G]}-k} = \sqrt{\dfrac{1}{k} \sum_{i=M^{[G]}}^{M^{[G]}-k} \left( \Delta_{i}^{[G]} - \mu_{\Delta_{M^{[G]}-k}^{{[G]}}} \right)^2},
\end{align}
where the $\mu_{\Delta_{M^{[G]}-k}^{[G]}}$ is the mean value of $\Delta_{i}^{[G]}$ for $i = M^{[G]} - k, ... M^{[G]}$ that is defined as
\begin{align}
\mu_{\Delta_{M^{[G]}-k}^{[G]}}=\dfrac{1}{k} \sum_{i=M^{[G]}}^{M^{[G]}-k} \Delta_{i}^{[G]}.
\end{align}

The standard deviation $\sigma_{M^{[G]}-k-1}$ takes into account all errors in previous steps. Intuitively, starting from the smallest noise global eigenvalue, the first signal global eigenvalue can be automatically found by comparing the relative prediction error and the standard deviation of the approximation errors in the previous steps. To this end, we consider the Prediction Error to Standard Deviation Ratio (PESDR)
\begin{align}
\text{PESDR}_{k}=  \frac{\delta_{M^{[G]}-k}}{\sigma_{{M^{[G]}-k}-1}}, \: k = 1, 2, ..., M^{[G]}-1.
\end{align}

To illustrate the behavior of the PESDR curve with respect to the profile of the global eigenvalues, we have constructed two random tensors with a known rank equal to five and dimensions 100$\times$350$\times$30 for the CP model \eqref{eq:X0}. The first tensor has uncorrelated columns in all factor matrices. The correlation of all factor matrices for the second tensor is defined by the vector $ \ma{r}_2 = [0.6 \; 0.7 \; 0.3]^\trans$ according to \cite{Edlin2015CholDec}. Gaussian noise is added to each tensor according to the CP model \eqref{eq:X} with SNR=-3 dB and SNR=11 dB for the uncorrelated and the correlated case, respectively. The SNR is defined as
\begin{align}
\text{SNR} = 10 \cdot \log_{10} \frac{\mathbb{E}\{{\fronorm{\ten{X}_0}}^2)\}}{\mathbb{E}\{({\fronorm{\ten{N}}}^2)\}} \text{, dB},
\label{eq:SNR}
\end{align}
where the $\mathbb{E}\{ \cdot \}$ denotes the expected value.
\begin{figure}[t]
	\vspace{-5mm}
	\centering
	\begin{subfigure}[htb]{1.0\linewidth}
		\includegraphics[width=\linewidth]{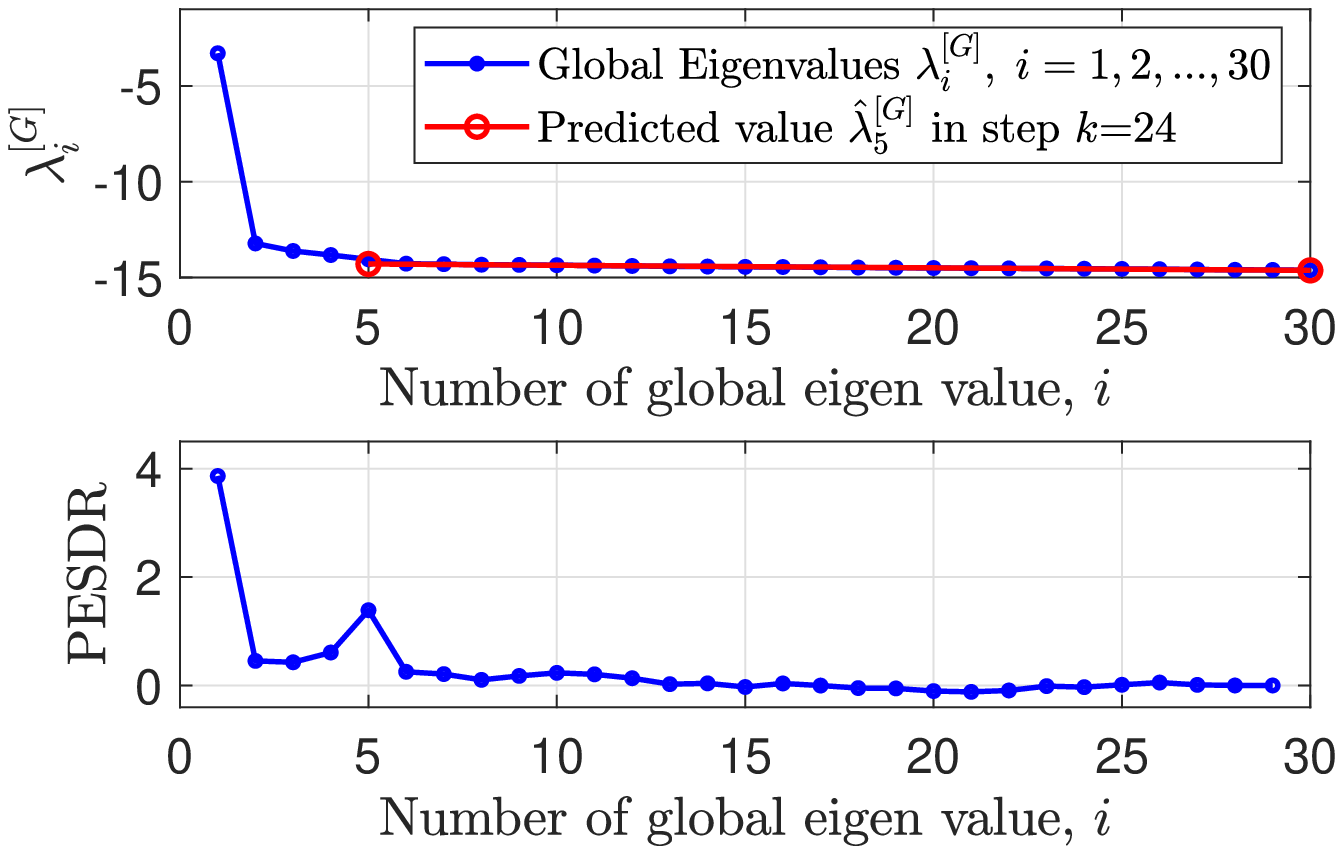}
		\caption{ }
	\end{subfigure}
	\begin{subfigure}[htb]{1.0\linewidth}
		\vspace{-1mm}
		\includegraphics[width=\linewidth]{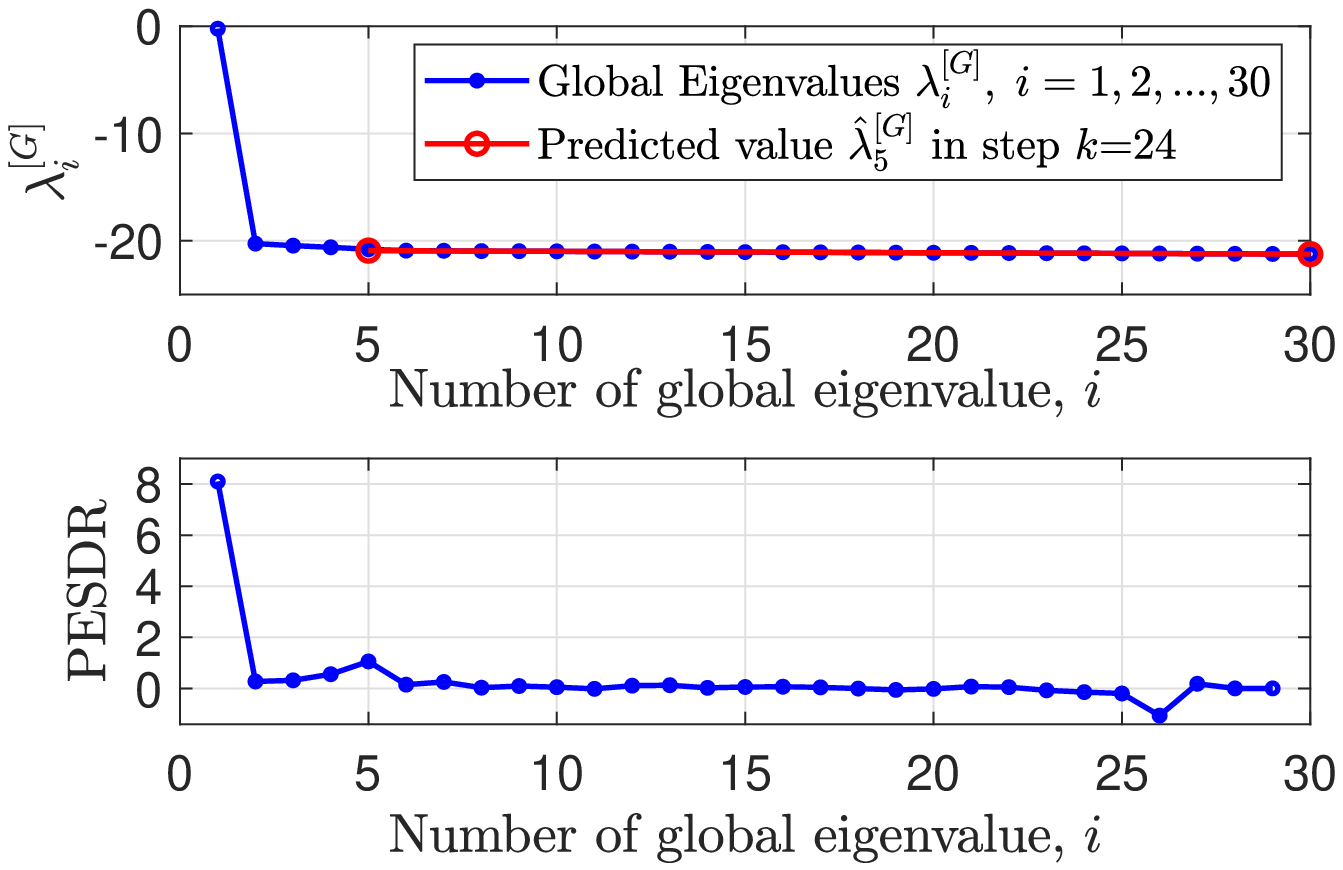}
		\caption{ }
	\end{subfigure}
	\caption{The profile of the global eigenvalues (top) and the Prediction Error to Standard Deviation Ratio (bottom) computed for a random tensor $\ten{X}~\in~\real^{100~\times~350~\times~30}$ with a known rank 5. (a) Uncorrelated case, SNR=-3 dB, (b) correlated case, SNR=11 dB}
	\vspace{-2.5mm}
	\label{fig:RelPESDR}
\end{figure}

Figure \ref{fig:RelPESDR} depicts the profile of the global eigenvalues and the computed PESDR coefficients for the constructed tensors. The straight line fits the profile of the noise global eigenvalues. It is seen that the first number $i$, where the PESDR curve exceeds a certain threshold $\rho$ corresponds to the rank of the noisy tensor. 

To formalize this criterion for the model order or the rank estimation, we can consider the following hypotheses:
\begin{align}
H_k^{(N)}: \lambda_{M^{[G]}-k}^{[G]} \; \text{is a noise global eigenvalue}\\
H_k^{(S)}: \lambda_{M^{[G]}-k}^{[G]} \; \text{is a signal global eigenvalue}.
\end{align}

In each step $k = 1, 2, ..., M^{[G]}-1$ or for each index $i=M^{[G]}-k$ the PESDR is compared to the threshold $\rho$ according to
\begin{align}
H_k^{(N)}: \text{PESDR}_{k}<\rho
\label{eq:HypoNoise}
\end{align}
\begin{align}
H_k^{(S)}: \text{PESDR}_{k}\geq\rho
\label{eq:HypoSignal}
\end{align}
The first index $k$ for which the hypothesis $H_k^{(N)}$ fails determines the estimated model order or the rank of the tensor and is denoted by $\hat{R}_\text{LaRGE} = M^{[G]}-k$.

\subsection{LaRGE with penalty function (LaRGE-PF)}

According to the global eigenvalues approach \cite{CostaHaardtRoemer2007}, \cite{Joao2011} the number of global eigenvalues that are used for model order estimation is equal to the smallest dimension of the tensor. In real-world applications, the smallest dimension can be relatively small compared to the rank. In the first steps of estimating the standard deviation of the approximation error  $\sigma_{M^{[G]}-k-1}$, the resulting estimate tends to zero. Hence, the PESDR can have outliers that exceed the threshold $\rho$. To avoid such a misclassification, we additionally ensure that  $\sigma_{M^{[G]}-k-1}$ exceeds a threshold $\epsilon$. In this paper, we set $\epsilon = 1.2\cdot10^{-3}$. 

Moreover, we also consider the slightly modified "PESDR with penalty function (PESDR-PF)" criterion that penalizes the standard deviation in the first steps
\begin{equation}
\begin{aligned}
\text{PESDR-PF}_k = \frac{\delta_{M^{[G]}-k}}{  \text{PF}(M^{[G]},k-1)\cdot \sigma_{{M^{[G]}-k}-1}} = \\
 =\frac{1}{\log_{10}(M^{[G]}-k-1)} \cdot \frac{\delta_{M^{[G]}-k}}{\sigma_{{M^{[G]}-k}-1} }.
\end{aligned}
\end{equation}

In the first estimation steps, the indices of the corresponding global eigenvalues $k$ are small and the argument of the logarithm $(M^{[G]}-k)$ in the denominator of the penalty function is relatively big. Therefore, the values of the penalty function (PF) have big weights. Hence, the denominator of the ratio becomes bigger and outliers are damped. As the number of steps increases, i.e., as the index $k$ increases, the value of the penalty function decreases and the damping effect is reduced. 

\subsection{Threshold determination}

The threshold $\rho$ is related to the statistical distribution of the absolute prediction error $\Delta_{i}^{[G]}$ or the relative prediction error $\delta_{i}^{[G]}$. This statistical distribution is not known for applications with measured real data. Therefore, numerical methods can instead be used for the threshold determination. 

In this paper, the threshold $\rho$ is obtained by the Monte Carlo simulations. For this purpose, we define the probability of false alarm or false positive $P_{\text{fp}}$, the probability of missing or false negative $P_{\text{fn}}$, and the probability of detection PoD.

The probability of false positive $P_{\text{fp}}$ is the probability that the LaRGE method falsely classifies a noise global eigenvalue as a signal global eigenvalue and is defined as
\begin{align}
P_{\text{fp}} = \text{Pr}\left( H_k^{(S)} \mid H_k^{(N)}  \right) =\text{Pr}\left(\hat{R}_{\text{LaRGE}}>R \mid R \right). 
\end{align}
A big value of this probability means that the rank or the model order can be overestimated. 

The probability of false negative  $P_{\text{fn}}$ expresses the probability that the proposed method mistakenly determines the model order smaller than the actual order. This probability is defined as
\begin{align}
P_{\text{fn}} = \text{Pr}\left( H_k^{(N)} \mid H_k^{(S)}  \right) =\text{Pr}\left(\hat{R}_{\text{LaRGE}}<R \mid R \right). 
\end{align}
The model order or the rank can be underestimated if the value of $P_{\text{fn}}$ is big.

The probability of detection is the probability of correct model order estimation and is considered as
\begin{align}
\text{PoD} = \text{Pr}\left( \hat{R}=R \right).
\end{align}

By considering a large number of trials we have obtained the functional dependency of the probabilities of false positive $P_{\text{fp}}(\rho)$, false negative $P_{\text{fn}}(\rho)$, and the probability of detection $\text{PoD}(\rho)$ as a function of the threshold $\rho$. 

\begin{figure}[t]
	\vspace{-5mm}
	\centering
	\includegraphics[width=1.05\linewidth]{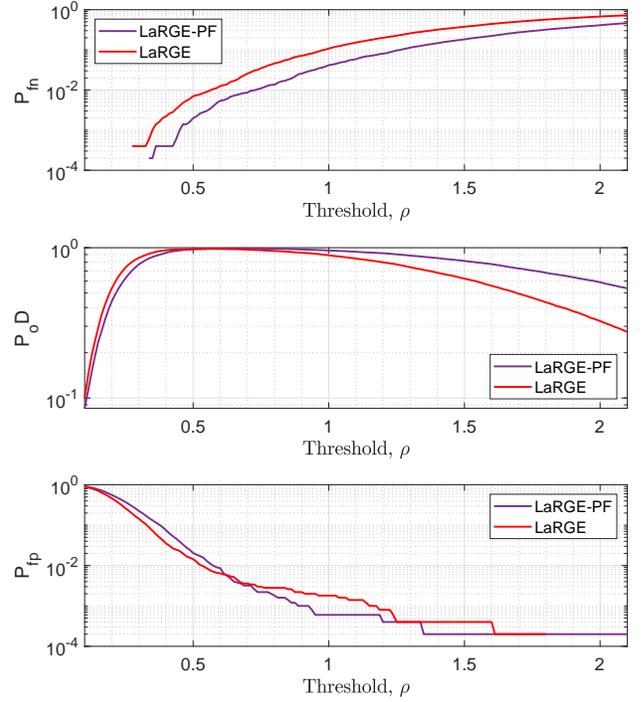} 
	\vspace{-10mm}
	\caption{The functional dependencies of $P_{\text{fn}}$ (top), PoD (middle), $P_{\text{fp}}$ (bottom) and the threshold $\rho$ for the noisy CP model, SNR=8 dB, $\ten{X}~\in~\real^{25~\times~30~\times~35}$, the true rank is $R=5$, 5000 trials.}
	\vspace{-2.5mm}
	\label{fig:prvsthr253035r5}
\end{figure}
To this end, we generate a data tensor $\ten{X}_0$ for the CP model \eqref{eq:X0}. All elements of the factor matrices are drawn from a zero-mean real-valued Gaussian distribution. Moreover, all columns of the factor matrices $\ma{F}_d,\;d=1,2,3,\dots D$ are not correlated. For all simulations, noiseless tensors of rank $ R=5 $ are constructed. The elements of the noise tensor $\ten{N}$ are drawn independently from a zero-mean Gaussian distribution with the variance $\sigma_n^2$ and the SNR that is defined as in \eqref{eq:SNR}.

For the first simulation, the size of the noisy tensor is set to $M_1=25, M_2=30, M_3=35$. The SNR is fixed at SNR=8 dB, and the value of the threshold $\rho$ varies in the range from 0.1 to 2.1. The curves of the probabilities $P_{\text{fp}}$, $P_{\text{fn}}$, and PoD of the LaRGE method obtained after 5000 Monte Carlo trials are shown in Figure \ref{fig:prvsthr253035r5}.

\begin{figure}[t]
	\vspace{-5mm}
	\centering
	\includegraphics[width=1.05\linewidth]{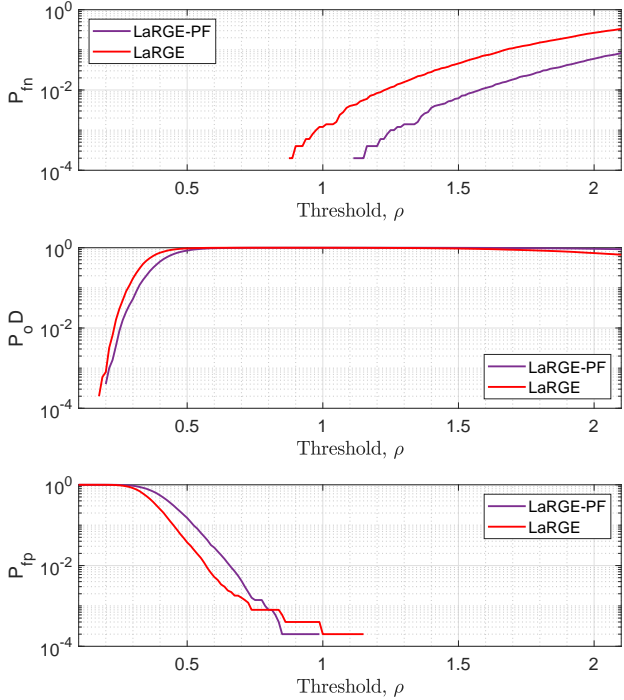} 
	\vspace{-10mm}
	\caption{The functional dependencies of $P_{\text{fn}}$ (top), PoD (middle), $P_{\text{fp}}$ (bottom) and the threshold $\rho$ for the noisy CP model, SNR=0 dB, $\ten{X}~\in~\real^{60~\times~100~\times~70}$, the true rank is $ R=5 $, 5000 trials.}
	\vspace{-2.9mm}
	\label{fig:prvsthr6010070r5}
\end{figure}

For the second simulation, the size of the noisy tensor is set to $M_1=60, M_2=100, M_3=70$ and SNR=0 dB. Figure \ref{fig:prvsthr6010070r5} shows the curves of the probabilities $P_{\text{fp}}$, $P_{\text{fn}}$, and PoD of the LaRGE method obtained after 5000 Monte Carlo trials for this scenario.

\begin{figure}[t]
	\vspace{-5mm}
	\centering
	\includegraphics[width=1.05\linewidth]{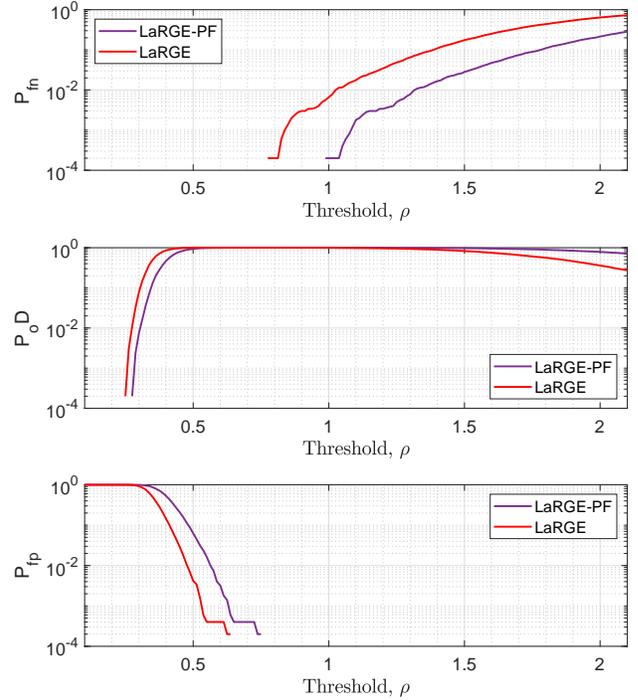}
	\vspace{-5mm}
	\caption{The functional dependencies of $P_{\text{fn}}$ (top), PoD (middle), $P_{\text{fp}}$ (bottom) and the threshold $\rho$ for the noisy CP model, SNR= -9 dB, $\ten{X}~\in~\real^{78~\times~1000~\times~102}$, the true rank is $ R=5 $, 5000 trials.}
	\vspace{-2.9mm}
	\label{fig:prvsthr781000102r5}
\end{figure}

For the third simulation, the size of the noisy tensor (SNR= -9 dB) is significantly increased and is set to $M_1=78, M_2=1000, M_3=102$. The resulting curves of the probabilities $P_{\text{fp}}$, $P_{\text{fn}}$, and PoD of the LaRGE method are depicted in Figure \ref{fig:prvsthr781000102r5}.

\begin{figure}[t]
	\vspace{-5mm}
	\centering
	\includegraphics[width=1.05\linewidth]{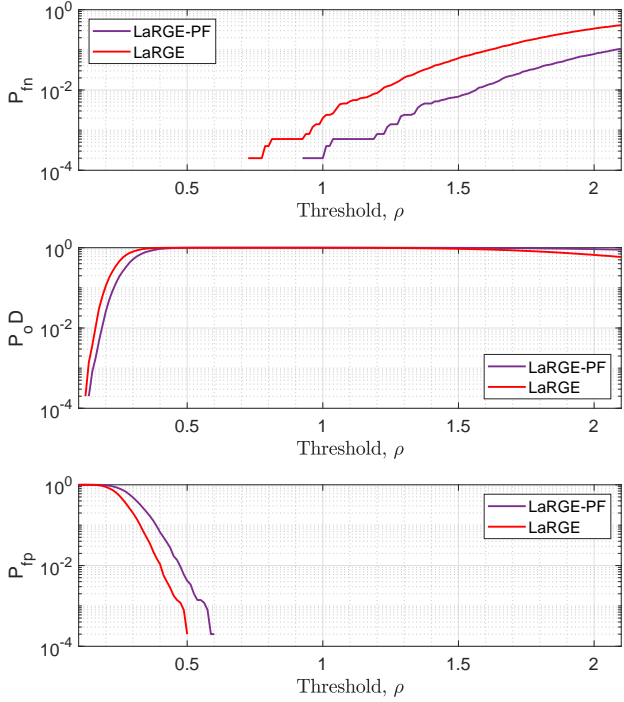}
	\vspace{-5mm}
	\caption{The functional dependencies of $P_{\text{fn}}$ (top), PoD (middle), $P_{\text{fp}}$ (bottom) and the threshold $\rho$ for the noisy CP model, SNR= -9 dB,  $\ten{X}~\in~\real^{60~\times~60~\times~60~\times~60}$, the true rank is $ R=5 $, 5000 trials.}
	\vspace{-2.9mm}
	\label{fig:prvsthr60606060r5}
\end{figure}

For the forth scenario, a four-dimensional noisy tensor (SNR= -9 dB) of size $M_1=60, M_2=60, M_3=60, M_4=60$ is generated. The obtained curves of the probabilities $P_{\text{fp}}$, $P_{\text{fn}}$, and PoD of the LaRGE method are shown in Figure \ref{fig:prvsthr60606060r5}.

Using the results of the Monte Carlo simulations, a numerical value for the threshold $\rho$ can be obtained by setting the probabilities $P_{\text{fp}}$ and $P_{\text{fn}}$ to an appropriate value. In this paper, we want to keep these probabilities below 1\% as used in \cite{Quinlan2006ModelOS} and \cite{CostaHaardtRoemer2007}. Simulations for small tensors show that the probability $P_{\text{fn}} = 0.01$ corresponds to the threshold $\rho=0.57$ (Figure \ref{fig:prvsthr253035r5}). Moreover, a smaller value of the false negative probability $P_{\text{fn}}$ of the LaRGE-PF method eventually provides a gain in terms of the correct detection probability PoD as shown in Figure \ref{fig:prvsthr253035r5}. 

On the other hand, for bigger tensors (Figure \ref{fig:prvsthr6010070r5}) the threshold $\rho=0.57$ corresponds to $P_{\text{fp}} = 0.01$.   
Furthermore, Figure \ref{fig:prvsthr6010070r5} shows that for an implementation of LaRGE-PF for tensors of size $60 \times 5100 \times 70$ the probabilities $P_{\text{fp}}$ and $P_{\text{fn}}$ are equal to zero, and hence, a correct detection probability PoD = 100$\%$ is obtained for the values of $\rho$ from 1 to 1.2. Both LaRGE and LaRGE-PF can provide a probability of correct detection PoD = $ 100 \%$ for tensors of larger sizes or dimensions  as shown in Figures \ref{fig:prvsthr781000102r5} and \ref{fig:prvsthr60606060r5}. 

For instance, LaRGE provides  PoD = $ 100 \%$ for the threshold $\rho$ in the range from 0.63 to 0.77 and LaRGE-PF in the range from 0.75 to 0.98 for tensors of size $78 \times 1000 \times 102$. For 4-dimensional tensors of size $60 \times 60 \times 60 \times 60$ a probability of correct detection PoD = $ 100 \%$ can be obtained for the threshold $\rho$ in the range from 0.5 to 0.75 using LaRGE and in the range from 0.6 to 0.92 using LaRGE-PF.

\section{Classical methods for multi-dimensional model order estimation}
\label{sec:Classical_methods}

To compare the performance of the proposed LaRGE method to the state-of-the-art, we use AIC and MDL  as well as their multi-dimensional extensions as a benchmark in this paper. 

The classical AIC and MDL methods are based on information theory and these methods have originally been applied to one dimensional data. According to \cite{WaxTail1985} the AIC criterion is given by

\begin{align}
\text{AIC}_{i}=-2{N^{[s]}(M^{[\sigma]}-i)}L(i) + 2i\left( 2M^{[\sigma]}-i \right),
\label{eq:AIC}
\end{align}
where $N^{[s]}$ and $M^{[\sigma]}$ are the number of snapshots and the number of the eigenvalues $\sigma_i$ of the covariance matrix $\ma{R_{xx}}$ of the observation $\ma{X}$, respectively. Note that the $L(i)$ in \eqref{eq:AIC} is the log-likelihood that represents the residual error and is defined as
\begin{align}
L(i)=\text{log}_{10}\left[ \dfrac{g(i)}{a(i)} \right].
\end{align}
The functions $g(i)$ and $a(i)$ are the geometric and arithmetic means of the eigenvalues $\sigma_i$ of the covariance matrix $\ma{R_{xx}}$, respectively. The second term in \eqref{eq:AIC} is a penalty for over-fitting. 

Contrary to the AIC criterion, the MDL criterion penalizes the over-fitting more strongly
\begin{equation}
\begin{aligned}
\text{MDL}_{i}=-{N^{[s]}(M^{[\sigma]}-i)}L(i) + \\
+ \dfrac{1}{2}i\left( 2M^{[\sigma]}-i \right)\text{log}_{10}\left( N^{[s]}\right).
\label{eq:MDL}
\end{aligned}
\end{equation}

Minimizing the following cost functions, the model order can be estimated as
\begin{align}
\hat{R}_{\text{AIC}}=\text{arg}\min\limits_{i} \left\lbrace \text{AIC}_{i} \right\rbrace,
\end{align}
\begin{align}
\hat{R}_{\text{MDL}}=\text{arg}\min\limits_{i} \left\lbrace \text{MDL}_{i} \right\rbrace.
\end{align}

In \cite{CostaHaardtRoemer2007} and \cite{Joao2011}, the authors have proposed an extension of the AIC \eqref{eq:AIC} and MDL \eqref{eq:MDL} methods to multidimensional data. This can be achieved by replacing the eigenvalues $\sigma_i$ of the covariance matrix $\ma{R_{xx}}$ by the global eigenvalues ${\lambda}_i^{[G]}$ defined in equation \eqref{eq:GlEigen}.
Replacing the eigenvalues of the covariance matrix by the global eigenvalues in \eqref{eq:AIC} - \eqref{eq:MDL}, with some straightforward manipulations, we obtain
\begin{align}
\text{AIC}_{i}^{[G]}=-2{N^{(s)}(M^{[G]}-i)}L^{[G]}(i) + 2i\left( 2M^{[G]}-i \right),
\label{eq:AIC_G}
\end{align}
\begin{equation}
\begin{aligned}
\text{MDL}_{i}^{[G]}=-{N^{(s)}(M^{[G]}-i)}L^{[G]}(i) +\\
+ \dfrac{1}{2}i\left( 2M^{[G]}-i \right)\text{log}_{10}\left( N^{(s)}\right),
\label{eq:MDL_G}
\end{aligned}
\end{equation}
where
\begin{align}
L^{[G]}(i)=\text{log} \left[ \dfrac { \left(  \prod\limits_{k=i+1}^{M^{[G]}} \lambda_{k}^{[G]} \right) ^{\dfrac{1}{M^{[G]}-i}} } {\dfrac{1}{M^{[G]}-i} \cdot \sum\limits_{k=i+1}^{{M^{[G]}}}\lambda_{k}^{[G]}} \right].
\end{align}

Finaly, the model order is estimated by the minimization of the costs functions
\begin{align}
\hat{R}_{N- \text{D AIC}}=\text{arg}\min\limits_{i} \left\lbrace \text{AIC}_{i}^{[G]} \right\rbrace,
\end{align}
\begin{align}
\hat{R}_{N- \text{D MDL}}=\text{arg}\min\limits_{i} \left\lbrace \text{MDL}_{i}^{[G]} \right\rbrace.
\end{align}


\section{Experimental results. Synthetic simulations}
\label{sec:Results}

In this section, we present the results of simulations to demonstrate the performance of the proposed LaRGE and LaRGE-PF methods in comparison with the classical AIC, MDL, $N$-D AIC, and $N$-D MDL methods. We set the threshold $\rho = 0.57$ for the LaRGE and LaRGE-PF methods and compare the probability of correct detection $\text{PoD}=\text{Pr}(\hat{R} = R)$ versus the SNR. 

A data tensor for the CP model \eqref{eq:X0} with factor matrices $ \ma{F}_d, d=1,2,3 \dots D$ is generated. The factor matrices contain elements drawn from a zero-mean Gaussian distribution with unit variance $\sigma_s^2$. 
Two different scenarios are used for the simulations with respect to the correlation of the columns in the factor matrices. In the first scenario, the correlation between the columns in the factor matrices $ \ma{F}_d, d=1,2,3 \dots D$ is parameterized by the vectors $ \ma{r}_1 = [0 \; 0 \; 0]^\trans$ and $ \ma{r}_1 = [0 \; 0 \; 0 \;0]^\trans$ for 3-D and 4-D tensors, respectively. The second scenario corresponds to the correlated case, and the correlation of the columns of factor matrices is parameterized by the vectors $ \ma{r}_2 = [0.6 \; 0.3 \; 0.9]^\trans$ and $ \ma{r}_2 = [0.6 \; 0.3 \; 0.9 \; 0.1]^\trans$ for 3-D and 4-D tensors, respectively \cite{ROEMER20132722}. The rank of the generated noiseless tensors is fixed at $R=5$. According to the model, \eqref{eq:X} the elements of the noise tensor $\ten{N}$ are drawn independently from a zero-mean Gaussian distribution with variance $\sigma_n^2$. 

\begin{figure}[t!]
	\centering
	\begin{subfigure}[htb]{1.0\linewidth}
		\includegraphics[width=\linewidth]{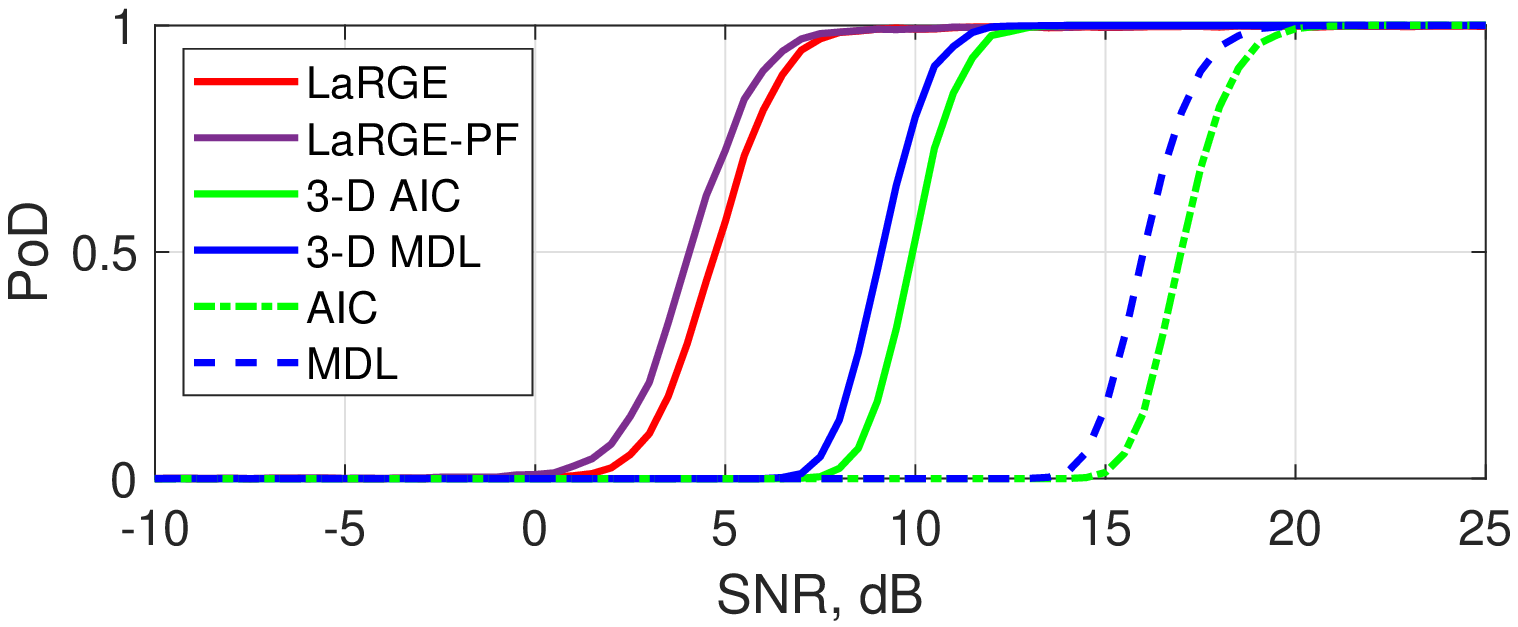}
	\end{subfigure}
	\begin{subfigure}[htb]{1.0\linewidth}
		\includegraphics[width=\linewidth]{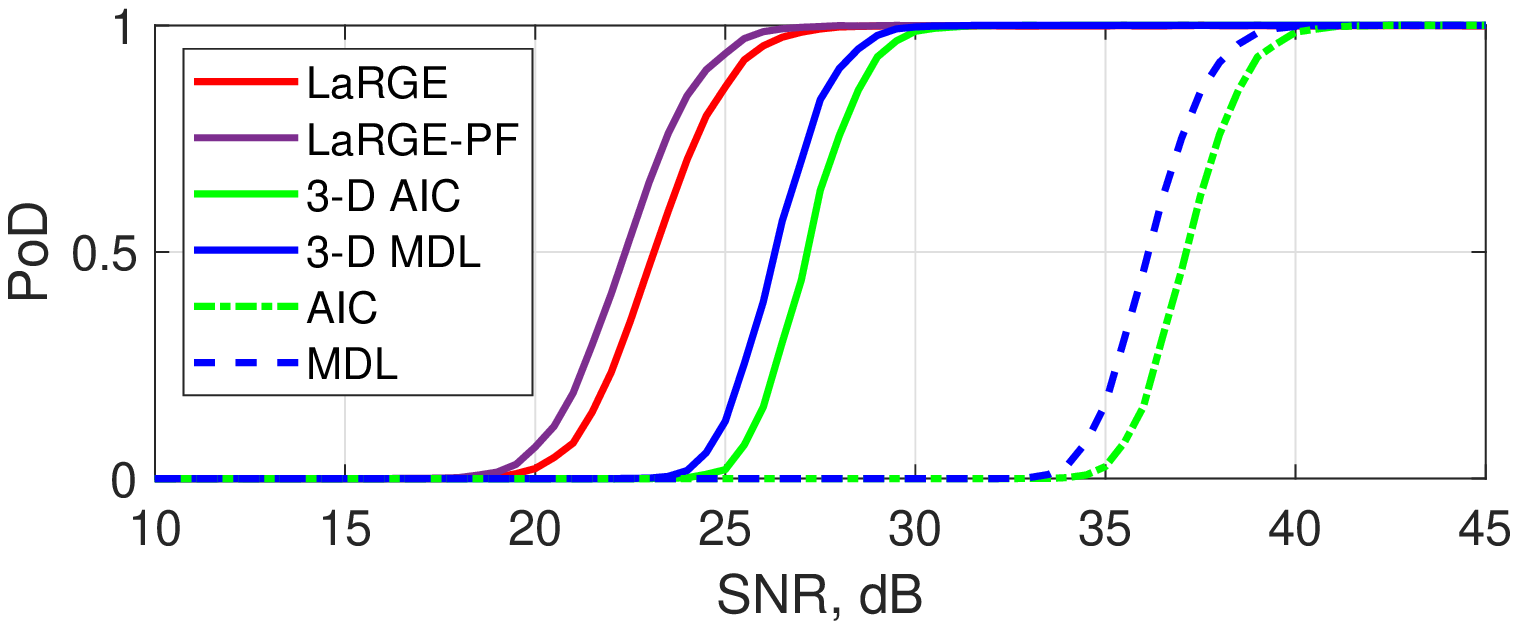}
	\end{subfigure}
	\caption{Probability of correct detection PoD versus the SNR for the noisy CP model, threshold $\rho = 0.57$, $\ten{X}~\in~\real^{25~\times~30~\times~35}$, the true rank is $ R=5 $, 5000 Monte Carlo trials, uncorrelated case (top), correlated case (bottom).}
	\vspace{-2.5mm}
	\label{fig:prvsSNR253035r5}
\end{figure}

In Figure \ref{fig:prvsSNR253035r5} we observe the performance of the classical methods AIC, MDL, $3$-D AIC, $3$-D MDL and the proposed LaRGE and LaRGE-PF methods for a tensor with the following dimensions $M_1=25, M_2=30, M_3=35$. The obtained results confirm the better performance of the $3$-D extension of AIC and MDL. Also, the proposed LaRGE and LaRGE-PF methods outperform the classical methods significantly.

\begin{figure}[t!]
	\centering
	\begin{subfigure}[htb]{1.0\linewidth}
		\includegraphics[width=\linewidth]{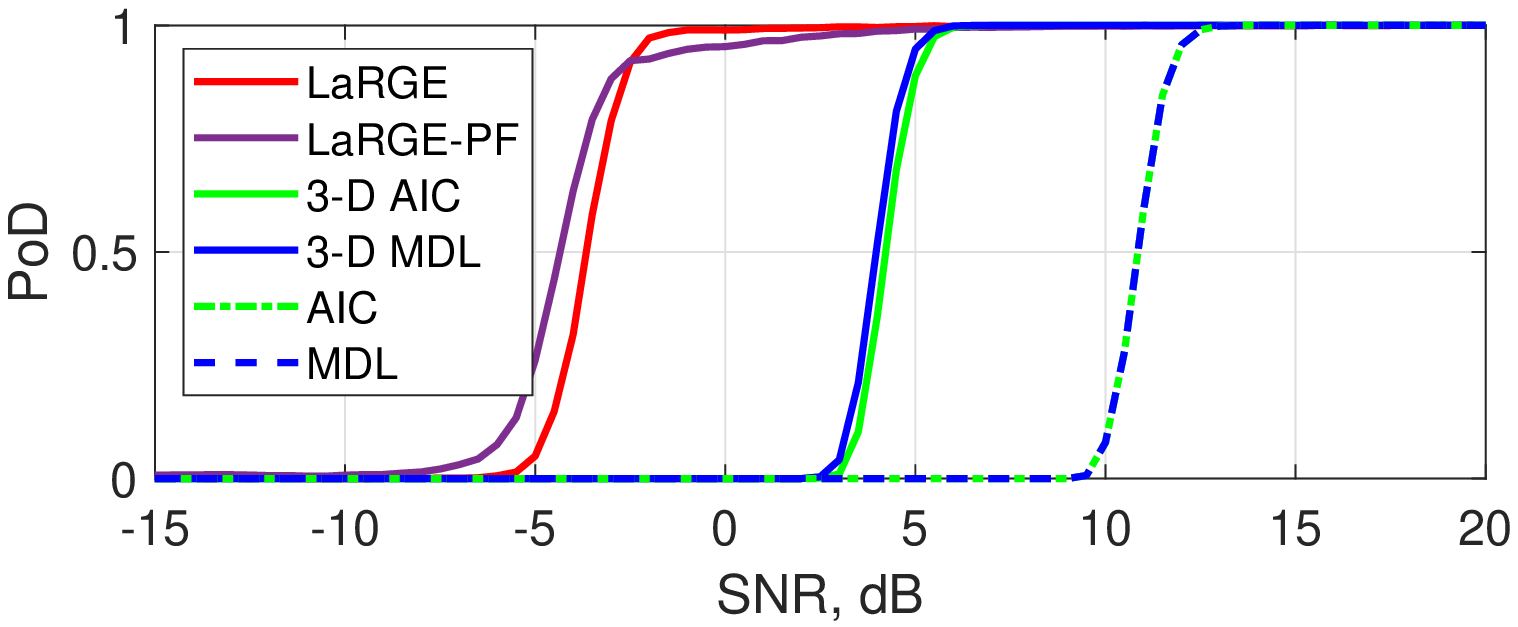}
	\end{subfigure}
	\begin{subfigure}[htb]{1.0\linewidth}
		\includegraphics[width=\linewidth]{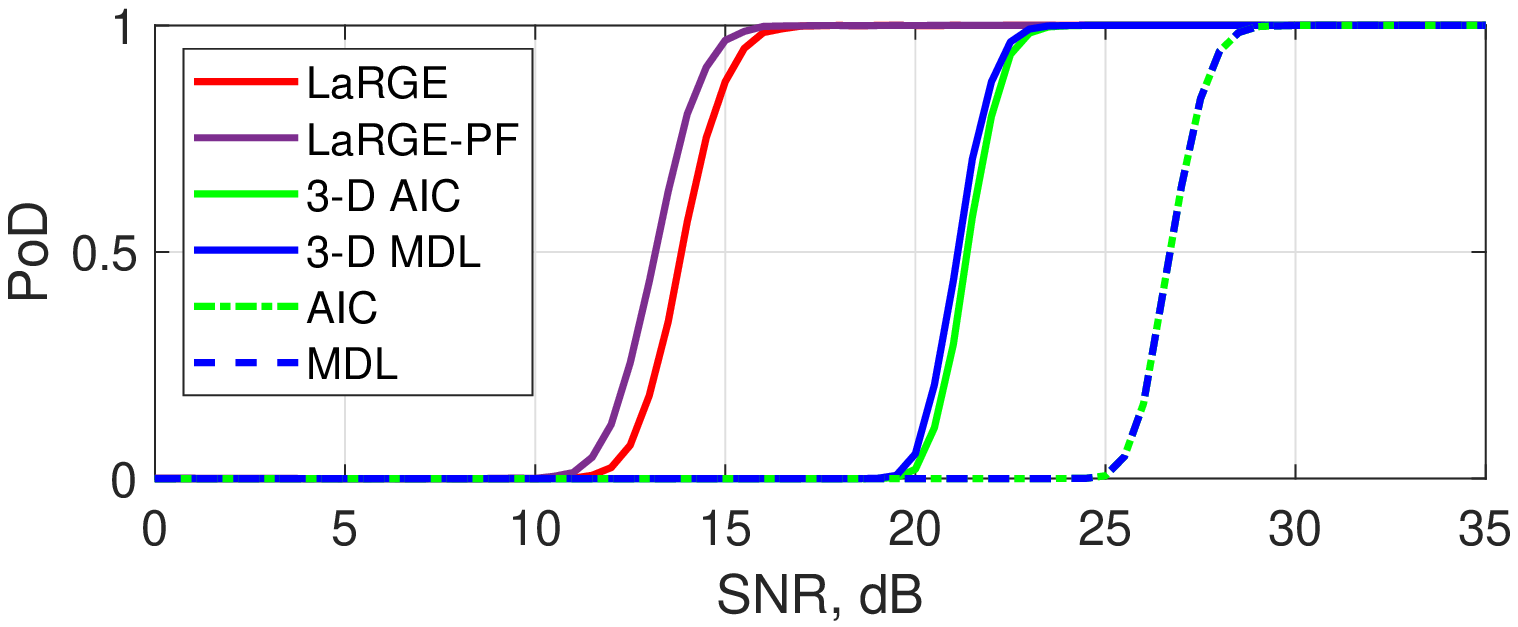}
	\end{subfigure}
	\caption{Probability of correct detection PoD versus the SNR for the noisy CP model, threshold $\rho = 0.57$, $\ten{X}~\in~\real^{60~\times~100~\times~70}$, the true rank is $ R=5 $, 5000 Monte Carlo trials, uncorrelated case (top), correlated case (bottom). }
	\vspace{-2.5mm}
	\label{fig:prvsSNR6010070r5}
\end{figure}

Similarly, in Figure \ref{fig:prvsSNR6010070r5} we evaluate the probability of correct detection of the AIC, MDL, $3$-D AIC, $3$-D MDL, LaRGE, and LaRGE-PF methods for a tensor with the following dimensions $M_1=60, M_2=100, M_3=70$. The LaRGE method shows a better performance in comparison with LaRGE-PF for SNRs up to 5 dB in the uncorrelated case.

\begin{figure}[t!]
	\centering
	\begin{subfigure}[htb]{1.0\linewidth}
		\includegraphics[width=\linewidth]{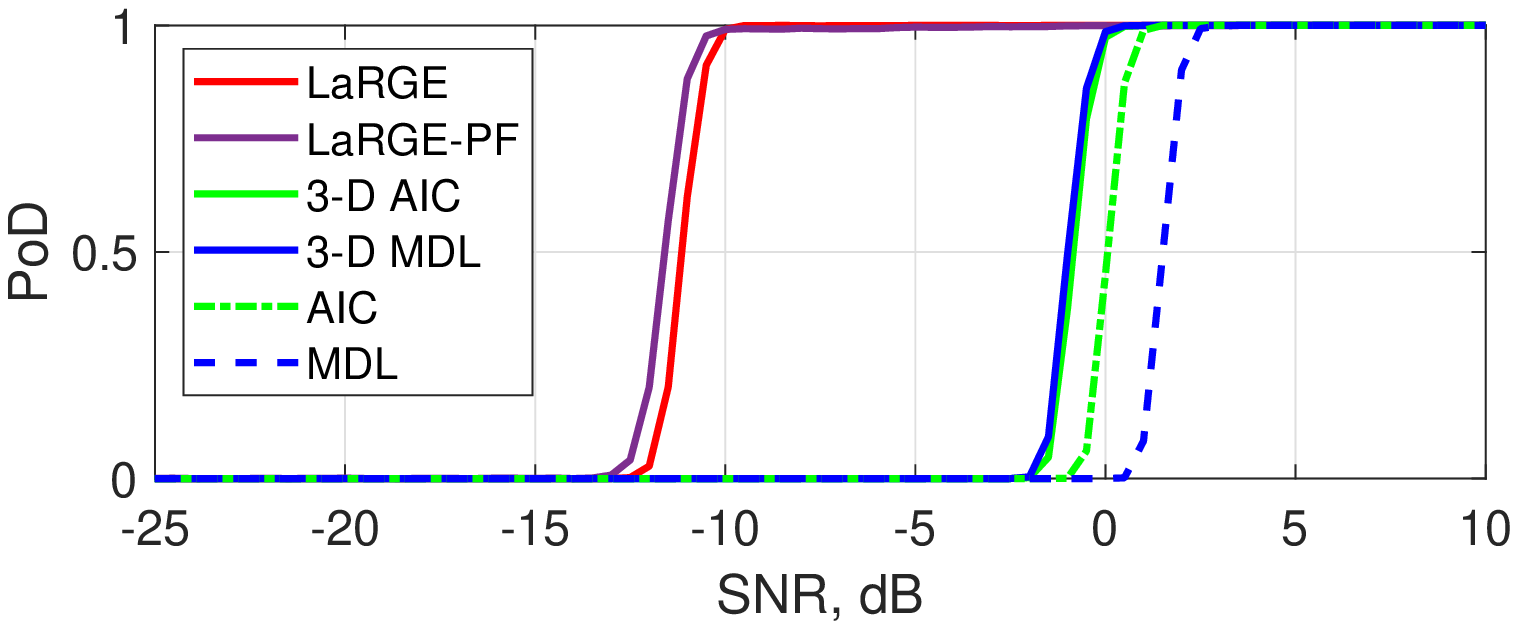}
	\end{subfigure}
	\begin{subfigure}[htb]{1.0\linewidth}
		\includegraphics[width=\linewidth]{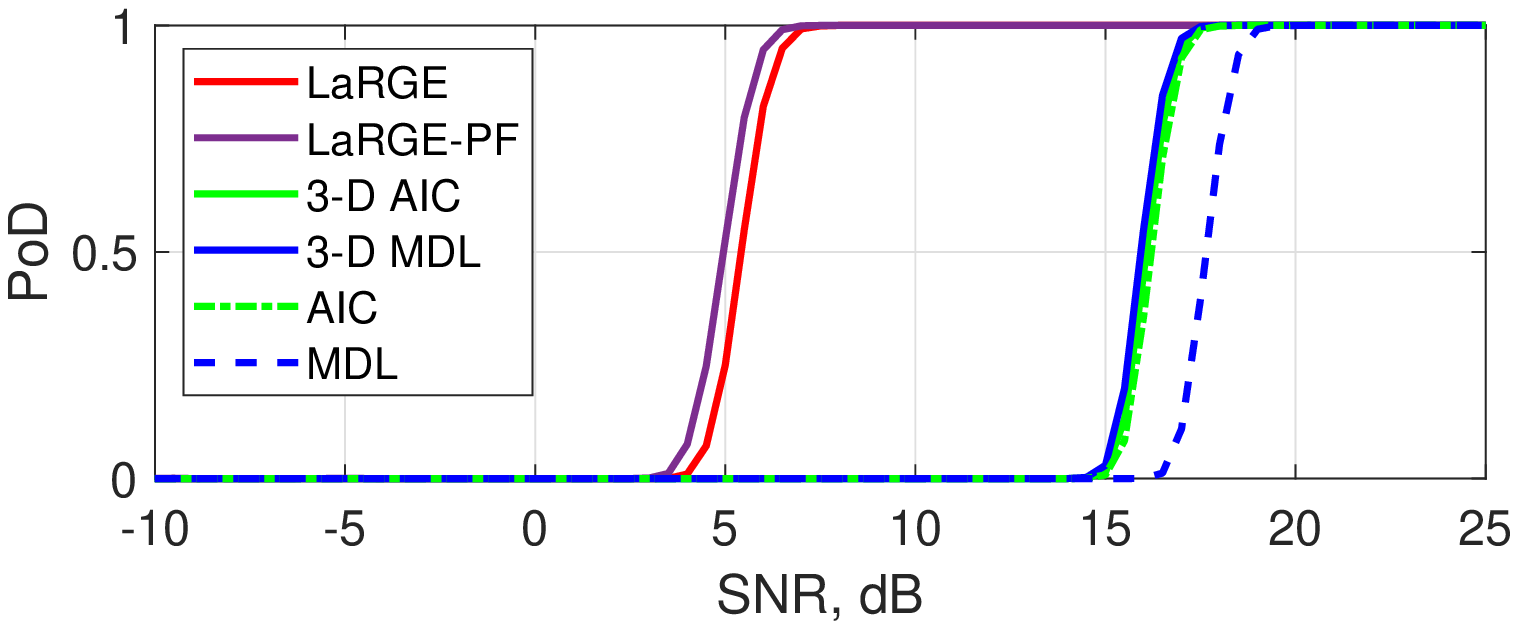}
	\end{subfigure}
	\caption{Probability of correct detection PoD versus the SNR for the noisy CP model, threshold $\rho = 0.57$, $\ten{X}~\in~\real^{78~\times~1000~\times~102}$, the true rank is $ R=5 $, 5000 Monte Carlo trials, uncorrelated case (top), correlated case (bottom). }
	\vspace{-2.5mm}
	\label{fig:prvsSNR781000102r5}
\end{figure}

Figure \ref{fig:prvsSNR781000102r5} depicts a scenario where the size of the tensor of rank $R=5$ has the dimensions $M_1=78, M_2=1000, M_3=102$. Again, the LaRGE and LaRGE-PF methods outperform AIC, MDL, $3$-D AIC, and $3$-D MDL. We can see that the $3$-D AIC, $3$-D MDL, AIC, and MDL methods show similar results. This effect can be explained as follows. 

For AIC and MDL, the $d$-mode singular values are used from the biggest dimension of the tensor. In our case, the largest number of $d$-mode singular values is equal to 1000 (for $d = 2$). However, for  $3$-D AIC and $3$-D MDL the number of global eigenvalues is limited by the smallest tensor dimension. In our case, this number is 78. Nevertheless, LaRGE and LaRGE-PF enhance the performance significantly.

\begin{figure}[t!]
	\centering
	\begin{subfigure}[htb]{1.0\linewidth}
		\includegraphics[width=\linewidth]{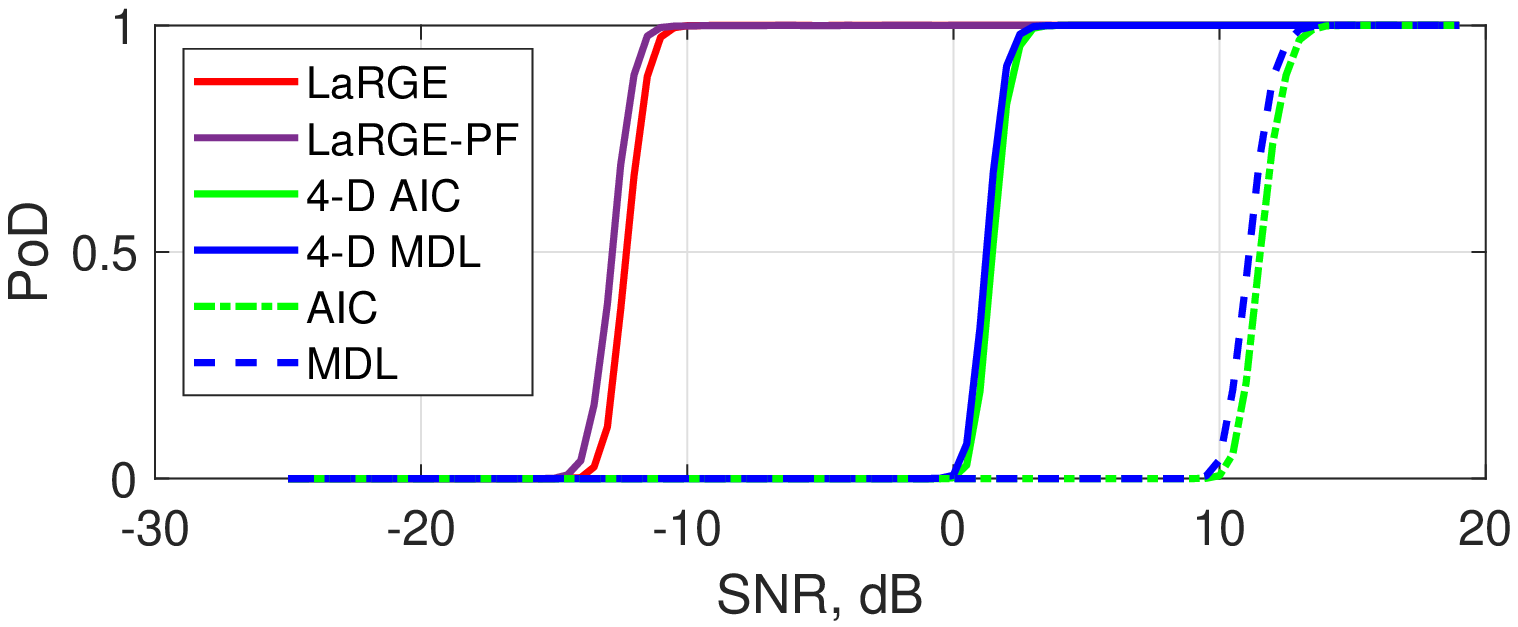}
	\end{subfigure}
	\begin{subfigure}[htb]{1.0\linewidth}
		\includegraphics[width=\linewidth]{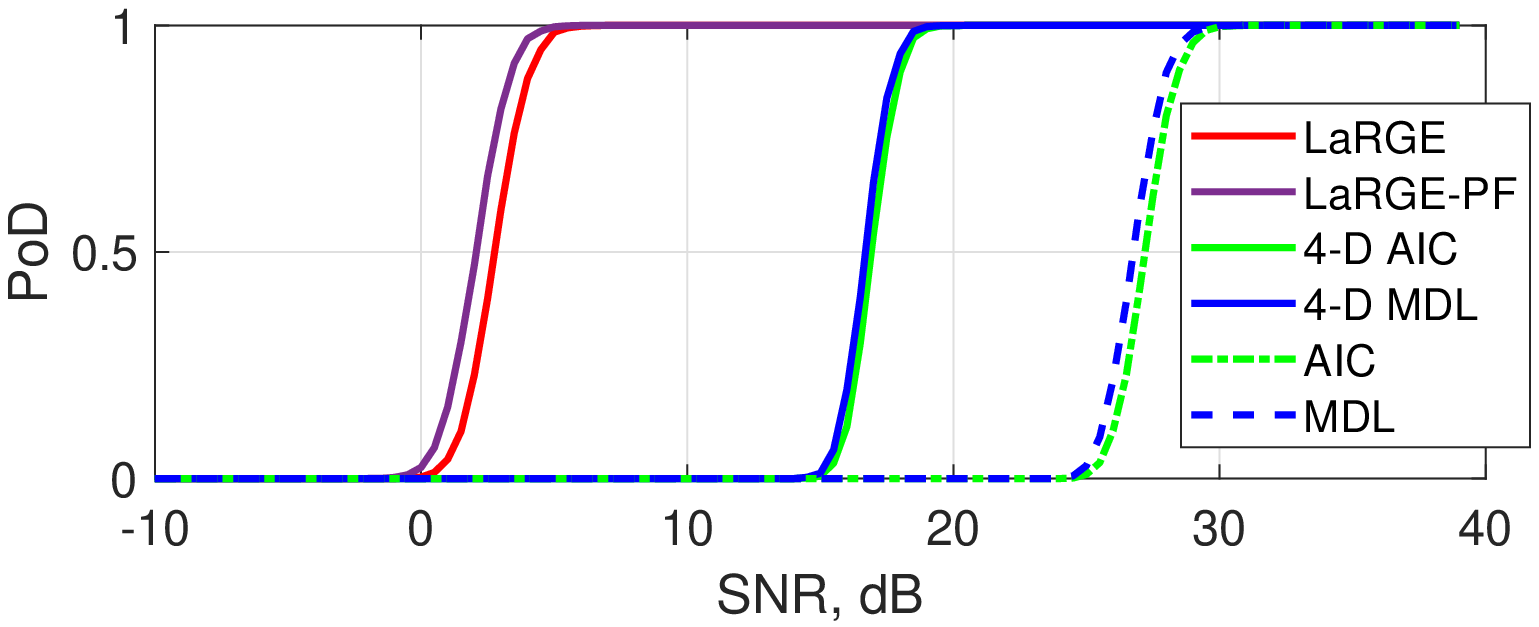}
	\end{subfigure}
	\caption{Probability of correct detection PoD versus the SNR for the noisy CP model, threshold $\rho = 0.57$, $\ten{X}~\in~\real^{60~\times~60~\times~60~\times~60}$, the true rank is $ R=5 $, 5000 Monte Carlo trials, uncorrelated case (top), correlated case (bottom).}
	\vspace{-2.5mm}
	\label{fig:prvsSNR60606060r5}
\end{figure}

We have also performed Monte Carlo simulations for a 4-D tensor scenario with dimensions $M_1=60, M_2=60, M_3=60, M_4=60$ and rank $R=5$. Figure \ref{fig:prvsSNR60606060r5} shows the results for this scenario. Again, we can clearly see that LaRGE and LaRGE-PF outperform the methods based on AIC and MDL by more than 12 dB. For small size tensors, the advantage of LaRGE-PF is more pronounced than for big 3-D tensors or 4-D tensors in comparison with the LaRGE method.

\section{Real biomedical signal processing}
\label{sec:ResultsRealData}
\subsection{Preprocessing and Decomposition}

The proposed LaRGE and LaRGE-PF methods presented in this paper are used to analyze measured EEG data. These data were recorded from twelve healthy volunteers, numbered 1 to 12 in this paper, at the Biomagnetic Center of the University Hospital in Jena, Germany, for the investigation of the Photic Driving (PD) effect \cite{SLJ+06},\cite{SaStKlJaSCHWi2016}. This effect occurs when the brain is stimulated by Intermittent Photic Stimulation (IPS). The PD effect is used to assess the effects of medications and for the diagnosis of several neurophysiological diseases like Alzheimer, schizophrenia, and some forms of epilepsy. 

The recorded EEG signals were filtered in the frequency range from 3 Hz to 40 Hz in a preprocessing step. The individual frequency of alpha rhythm $f_\alpha$ were determined for each volunteer prior to the main experiment. The resulting alpha frequencies for volunteers 1 to 12 are, in this order, 9.6, 10.7, 10.4, 10.8, 10.7, 10.8, 7.5, 10.8, 11.0, 10.7, 12.2, and 10.3 Hz. To investigate the PD effect, twenty stimulation frequencies $f_{stim}=$ [0.40 0.45 0.50 0.55 0.60 0.70 0.80 0.90 0.95 1.00 1.05 1.10 1.30 1.60 1.90 1.95 2.00 2.05 2.10 2.30] $\cdot f_\alpha$ were generated. Each stimulation of the particular frequency was presented in 30 stimulation trains. Each train consisted of 40 periods with pulse/cycle duration of 0.5. Between each train and frequency block there were resting periods of 4 seconds and 30 seconds, respectively.

To investigate the proposed LaRGE method for the model order estimation, two different cases are considered in this paper. 3-D tensors are constructed for the first case. To this end, the Fast Fourier Transform is used for obtaining the frequency distribution of the EEG signals. As a result, we have different complex tensors with dimensions channels $\times$ frequency $\times$ trains for each stimulation frequency and volunteer. The frequency and train dimensions have fixed sizes: 281 and 30, respectively. The size of the channel dimension varies, because a small number of non-functional EEG channels are excluded in the preprocessing step. The resulting 3-D tensors have dimensions channels$\times$frequency$\times$train.

For the second case that is considered in this paper, the wavelet decomposition is used to construct 4-D tensors. The second and third dimensions of these tensors represent the frequency-time distribution obtained after the complex Morlet decomposition. The wavelet coefficients between 3.003 Hz (1000/333 Hz) and 15.15 Hz (1000/66 Hz) are selected for the analysis. This frequency range spans the frequencies of the theta and alpha rhythms \cite{NaskLauKorHauHaa2020}. The size of the time dimension varies from 5 seconds up to 20 seconds depending on the stimulation frequency. The resulting 4-D tensors have dimensions channels $\times$frequency$\times$time$\times$train.

In the first step of the model order estimation, the HOSVD of the constructed tensors is computed. Next, the global eigenvalues are obtained and the model order is estimated using the LaRGE method, the LaRGE method with penalty function (LaRGE-PF), and classical AIC and MDL methods ($N$-D AIC, $N$-D MDL, AIC, and MDL). To obtain the factor matrices of the dominant components, the CP decomposition of each tensor is computed via SECSI \cite{ROEMER20132722} in the 3-D case and via SECSI-GU \cite{RoeSchrHaardt2012} in the 4-D case using the estimated rank of LaRGE and LaRGE-PF. The threshold $\rho = 0.57$ is fixed for both LaRGE and LaRGE-PF to ensure that the probability $P_{\text{fn}}$ is not greater than 0.01. To eliminate the scaling ambiguity, the loading factors of the components are computed if the rank is more than one according to
\begin{align}
\lambda_r = \prod_{d=1}^{D}\fronorm{\ma{F}(d,r)}, r=1 \dots \hat{R},
\label{eq:WeightFact}
\end{align}
where $\hat{R}$ is the estimated rank and, $D$ is the number of tensor dimensions.

In the rest of this section, we present some results that have been obtained during the analysis of 3-D and 4-D tensors. In the following figures, the profiles of global eigenvalues and the PESDR curves are shown. Moreover, the channel, frequency, and train signatures are presented for the 3-D case, and the channel, frequency, time, and train signatures are presented for the 4-D case. All components of the factor matrices are sorted in the order of decreasing loading factors computed according to (\ref{eq:WeightFact}).

\subsection{Processing of 3-D tensors}

\begin{figure}[t!]
	\centering
	\includegraphics[width=1.0\linewidth]{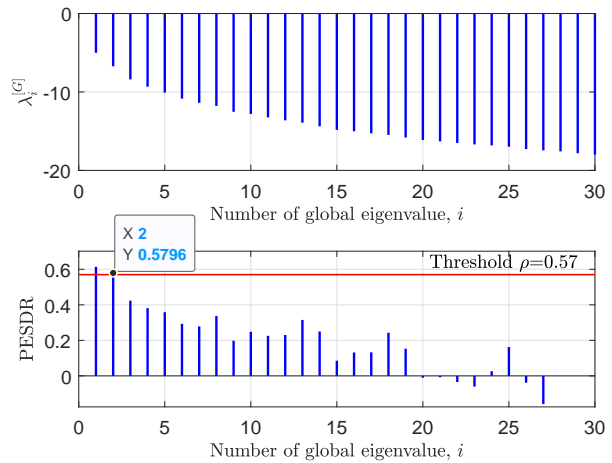} %
	\caption{The profile of the global eigenvalues (top) and Prediction Error to Standard Deviation Ratio (bottom) computed for the tensor with real EEG data $\ten{X}~\in~\compl^{103~\times~281~\times~30}$ of volunteer 4 at stimulation frequency 0.5$\cdot f_\alpha$.}
	\vspace{-2.5mm}
	\label{fig:3D_EEG_GEwPESDR_Vol_4_st_freq_3_R_2}
\end{figure}

\begin{figure}[t!]
	\centering
	\includegraphics[width=1.08\linewidth]{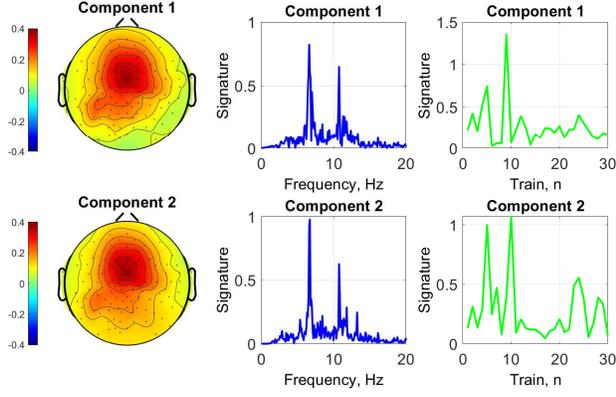} %
	\caption{Channel, frequency, and train signature for volunteer 4 and stimulation frequency 0.5$\cdot f_\alpha$ after the CP decomposition of the tensor with real EEG data $\ten{X}~\in~\compl^{103~\times~281~\times~30}$ and estimated rank 2. The loading factors of the components are $\lambda_1 = 0.2941$, $\lambda_2 = 0.2382$. }
	\vspace{-2.5mm}
	\label{fig:3D_EEG_FactMatr_Vol_4_st_freq_3_R_2}
\end{figure}

Figure \ref{fig:3D_EEG_GEwPESDR_Vol_4_st_freq_3_R_2} depicts the profile of the global eigenvalues and the PESDR coefficients for the tensor of volunteer 4 at stimulation frequency $0.5 \cdot f_\alpha$. The LaRGE method estimates the rank 2 for this case. 
Figure \ref{fig:3D_EEG_FactMatr_Vol_4_st_freq_3_R_2} depicts the dominant components of the factor matrices obtained via the CP decomposition. 
During the whole stimulation, the theta rhythm dominates which is evidenced by the distributions of both components in the spatial and frequency domains that are presented in Figure \ref{fig:3D_EEG_FactMatr_Vol_4_st_freq_3_R_2}. The other state-of-the-art model order estimation schemes considered in this paper have estimated the rank as follow $\hat R_{\rm {LaRGE-PF}}= 3$, $\hat R_{\rm {3-D AIC}}= 10$, $\hat R_{\rm {3-D MDL}}= 13$, $\hat R_{\rm {AIC}}= 49$, $\hat R_{\rm {MDL}}= 43$.

\begin{figure}[t!]
	\centering
	\includegraphics[width=1.0\linewidth]{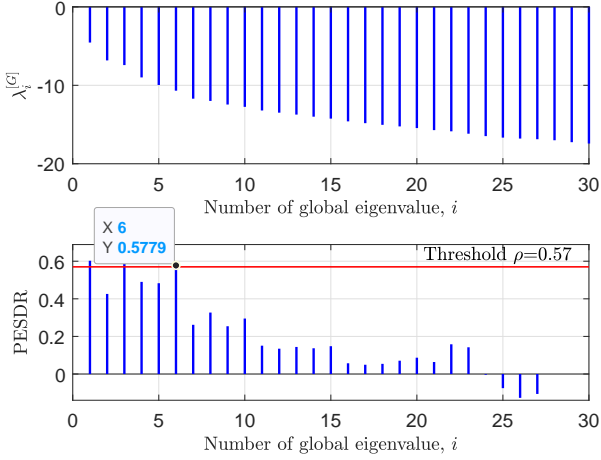} %
	\caption{The profile of the global eigenvalues (top) and Prediction Error to Standard Deviation Ratio (bottom) computed for the tensor with real EEG data $\ten{X}~\in~\compl^{102 \times 281 \times 30}$ of volunteer 11 at stimulation frequency 0.9$\cdot f_\alpha$.}
	\vspace{-2.5mm}
	\label{fig:3D_EEG_GEwPESDR_Vol_11_st_freq_8_R_6}
\end{figure}

\begin{figure}[t!]
	\centering
	\includegraphics[width=1.08\linewidth]{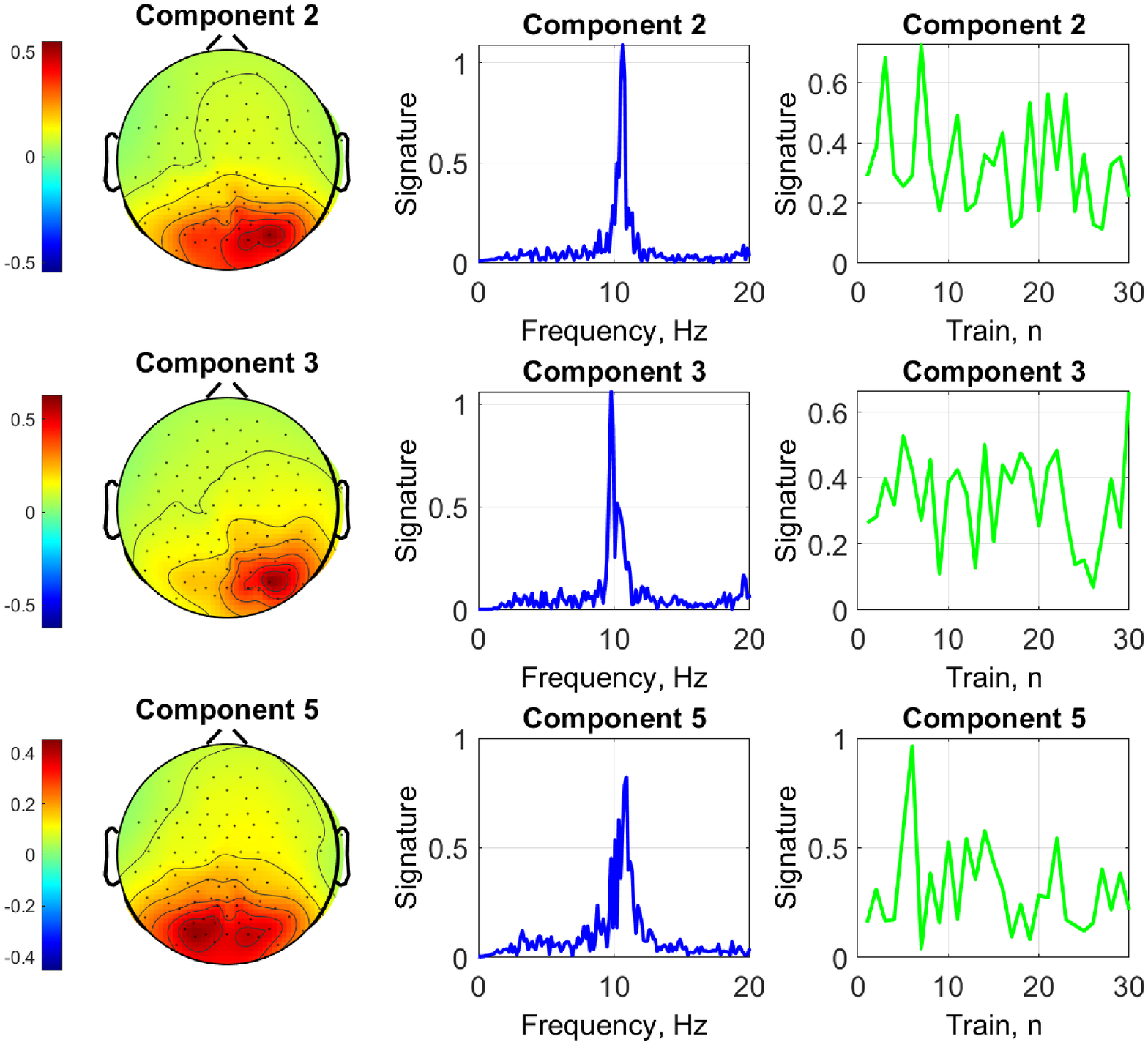} %
	\centering
	\includegraphics[width=1.08\linewidth]{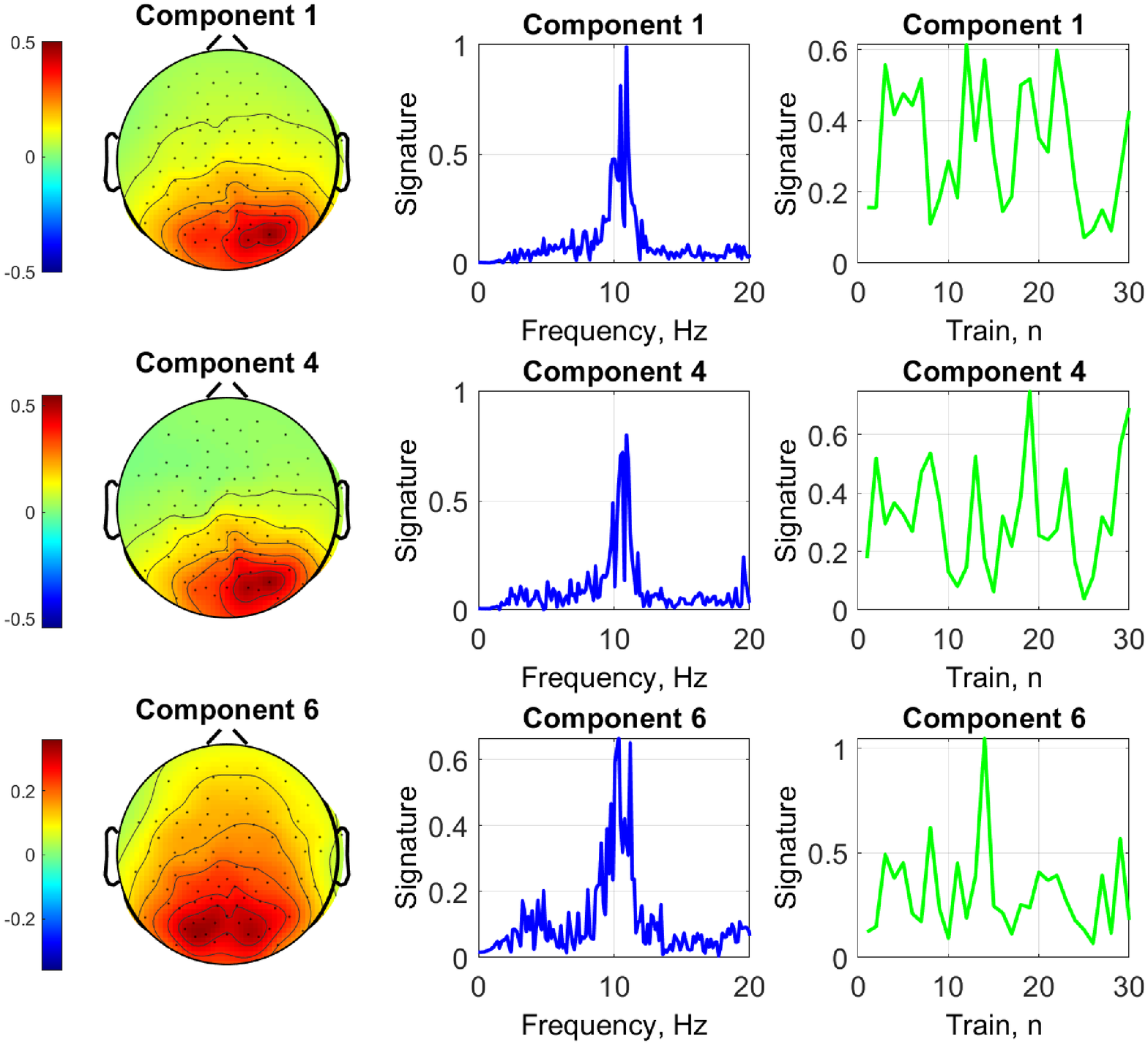} %
	\caption{Channel, frequency, and train signature for volunteer 11 and stimulation frequency 0.9$\cdot f_\alpha$ after the CP decomposition of the tensor with real EEG data $\ten{X}~\in~\compl^{102 \times 281 \times 30}$ and estimated rank 6. The loading factors of the components are $\lambda_2 = 0.2996$, $\lambda_3 = 0.2746$, $\lambda_5 = 0.2729$, $\lambda_1 = 0.2354$, $\lambda_4 = 0.2303$, $\lambda_6 = 0.2260$.} 
	\vspace{-2.5mm}
	\label{fig:3D_EEG_FactMatr_Vol_11_st_freq_8_R_3}
\end{figure}

The results of processing the EEG signals that were recorded from volunteer 11 at stimulation frequency $0.9 \cdot f_\alpha$ are presented in Figure \ref{fig:3D_EEG_GEwPESDR_Vol_11_st_freq_8_R_6} and Figure \ref{fig:3D_EEG_FactMatr_Vol_11_st_freq_8_R_3}. 

It is seen that the slope of the profile of the global eigenvalues changes between the sixth and the seventh global eigenvalue (see difference between 6th and 7th PESDR coefficient in Figure \ref{fig:3D_EEG_GEwPESDR_Vol_11_st_freq_8_R_6}). 
Therefore, a change of the slope of the global eigenvalues can be detected using the PESDR reliably to separate the signal and noise global eigenvalues.



The stimulation frequency $f_{\rm {stim}} = 0.90 \cdot f_\alpha$ is close to the individual alpha frequency. 
Note the variability in the train dimension across the six components which indicates variability in the individual processing of each stimulus sequence (Fig. \ref{fig:3D_EEG_FactMatr_Vol_11_st_freq_8_R_3}). The slightly variable topographies and frequency signatures for the 6 components also hint at a variable stimulation response possibly involving slightly different generator structures. The relatively high model order of six supports this interpretation.

\subsection{Processing of 4-D tensors}

\begin{figure}[t!]
	\centering
	\includegraphics[width=1.0\linewidth]{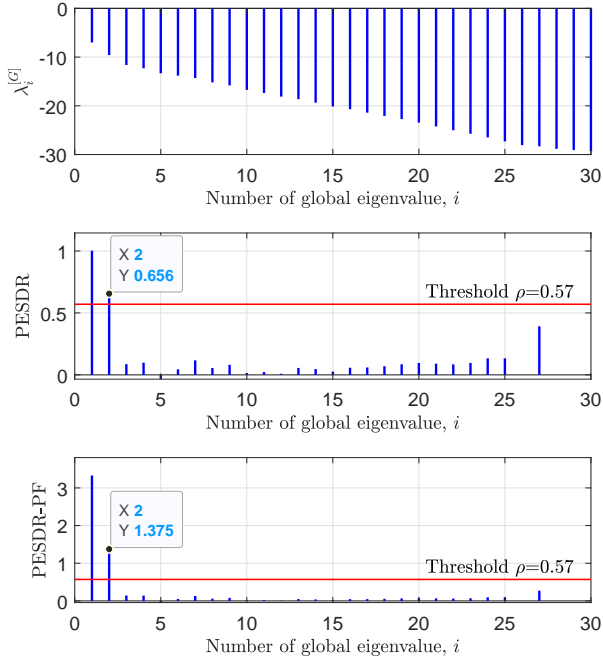} %
	\caption{The profile of the global eigenvalues (top), PESDR curve (middle) and PESDR-PF curve (bottom) computed for the tensor with real EEG data $\ten{X}~\in~\compl^{268 \times 838 \times 105 \times 30}$ of volunteer 1 at stimulation frequency 0.40$\cdot f_\alpha$.}
	\vspace{-2.5mm}
	\label{fig:4D_EEG_GEwPESDR_Vol_1_st_freq_1_R_2}
\end{figure}

\begin{figure}[t!]
	\centering
	\includegraphics[width=1.1\linewidth]{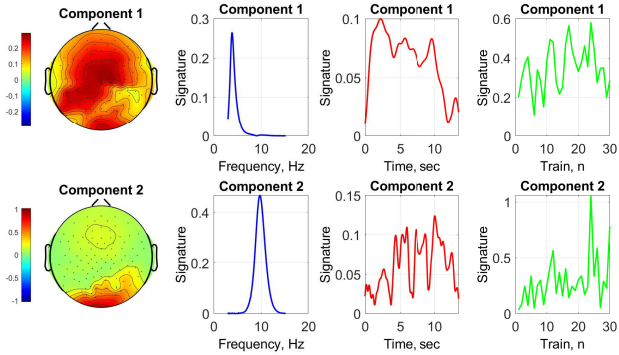} %
	\caption{Channel, frequency, time and train signature for volunteer 1 and stimulation frequency 0.40$\cdot f_\alpha$ after the CP decomposition of the tensor with real EEG data $\ten{X}~\in~\compl^{268 \times 838 \times 105 \times 30}$ and estimated rank 2. The loading factors of the components are $\lambda_1 = 0.1809$ $\lambda_2 = 0.1037$. }
	\vspace{-2.5mm}
	\label{fig:4D_EEG_FactMatr_Vol_1_st_freq_1_R_2}
\end{figure}

The results of processing a 4-D tensor with dimensions channel$\times$frequency$\times$time$\times$train for volunteer 1 are depicted in Figures \ref{fig:4D_EEG_GEwPESDR_Vol_1_st_freq_1_R_2} and \ref{fig:4D_EEG_FactMatr_Vol_1_st_freq_1_R_2}. The LaRGE method reliably estimates the model order as 2. Moreover, the bottom plot of Figure \ref{fig:4D_EEG_GEwPESDR_Vol_1_st_freq_1_R_2} shows the PESDR curve that is obtained according to LaRGE-PF. The suppression effect of the penalty function is clearly seen in this figure.

The dominant components of the factor matrices after the tensor decomposition show response in the alpha and theta frequencies. The alpha rhythm, as expected, is observed in the occipital region of the brain while the theta rhythm 
is dominant in the central region. Moreover, the theta rhythm has a resonance effect that is visible in the frequency and time dimensions in Figure \ref{fig:4D_EEG_FactMatr_Vol_1_st_freq_1_R_2} for component one. 
The other state-of-the-art model order estimation schemes considered in this paper estimate the model order of this tensor as $\hat R_{\rm {4-D AIC}}= 24$, $\hat R_{\rm {4-D MDL}}= 25$, $\hat R_{\rm {AIC}}= 205$, $\hat R_{\rm {MDL}}= 203$.


\subsection{Special cases}

\begin{figure}[t!]
	\centering
	\includegraphics[width=1.0\linewidth]{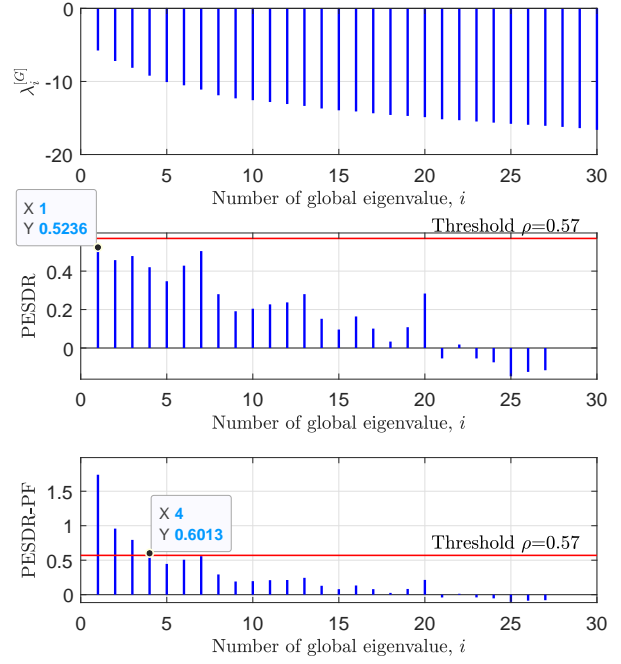} %
	\caption{The profile of the global eigenvalues (top), PESDR curve (middle) and PESDR-PF curve (bottom) computed for the tensor with real EEG data $\ten{X}~\in~\compl^{106 \times 281 \times 30}$ of volunteer 8 at stimulation frequency 0.50$\cdot f_\alpha$.}
	\vspace{-2.5mm}
	\label{fig:3D_EEG_GEwPESDR_Vol_8_st_freq_3_R_4}
\end{figure}

\begin{figure}[t!]
	\centering
	\includegraphics[width=1.1\linewidth]{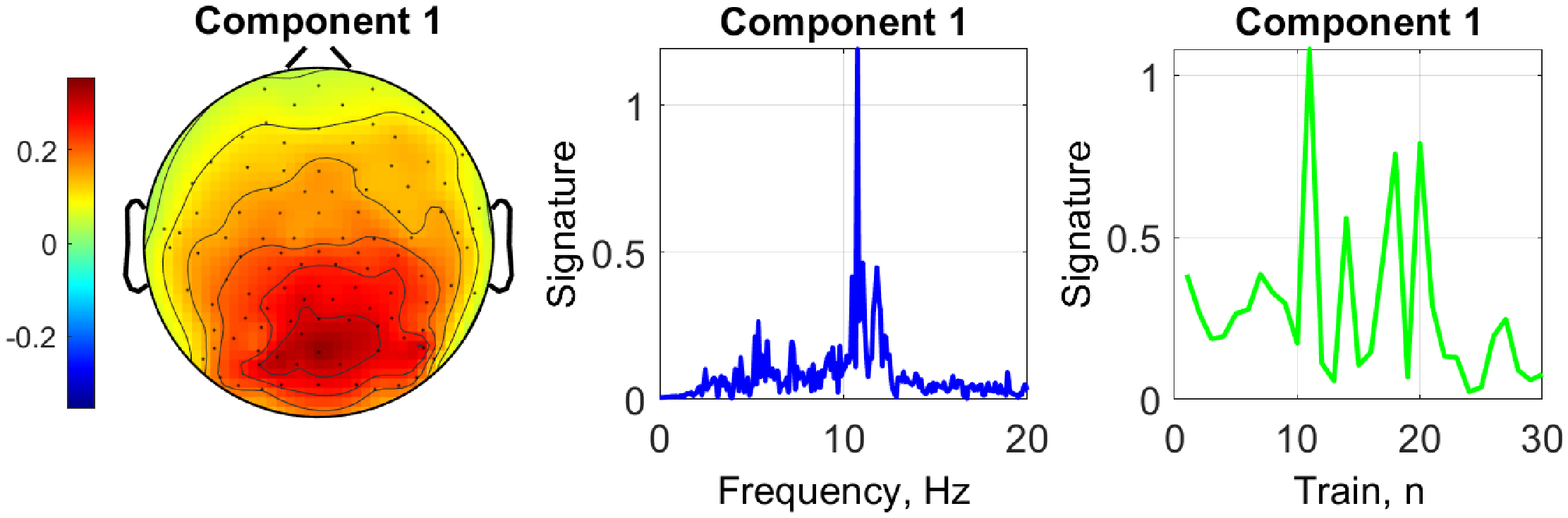} %
	\caption{Channel, frequency, time and train signature for volunteer 8 and stimulation frequency 0.50$\cdot f_\alpha$ after the CP decomposition of the tensor with real EEG data $\ten{X}~\in~\compl^{106 \times 281 \times 30}$ and estimated rank 1. }
	\vspace{-2.5mm}
	\label{fig:3D_EEG_FactMatr_Vol_8_st_freq_3_R_1}
\end{figure}

\begin{figure}[t!]
	\centering
	\includegraphics[width=1.08\linewidth]{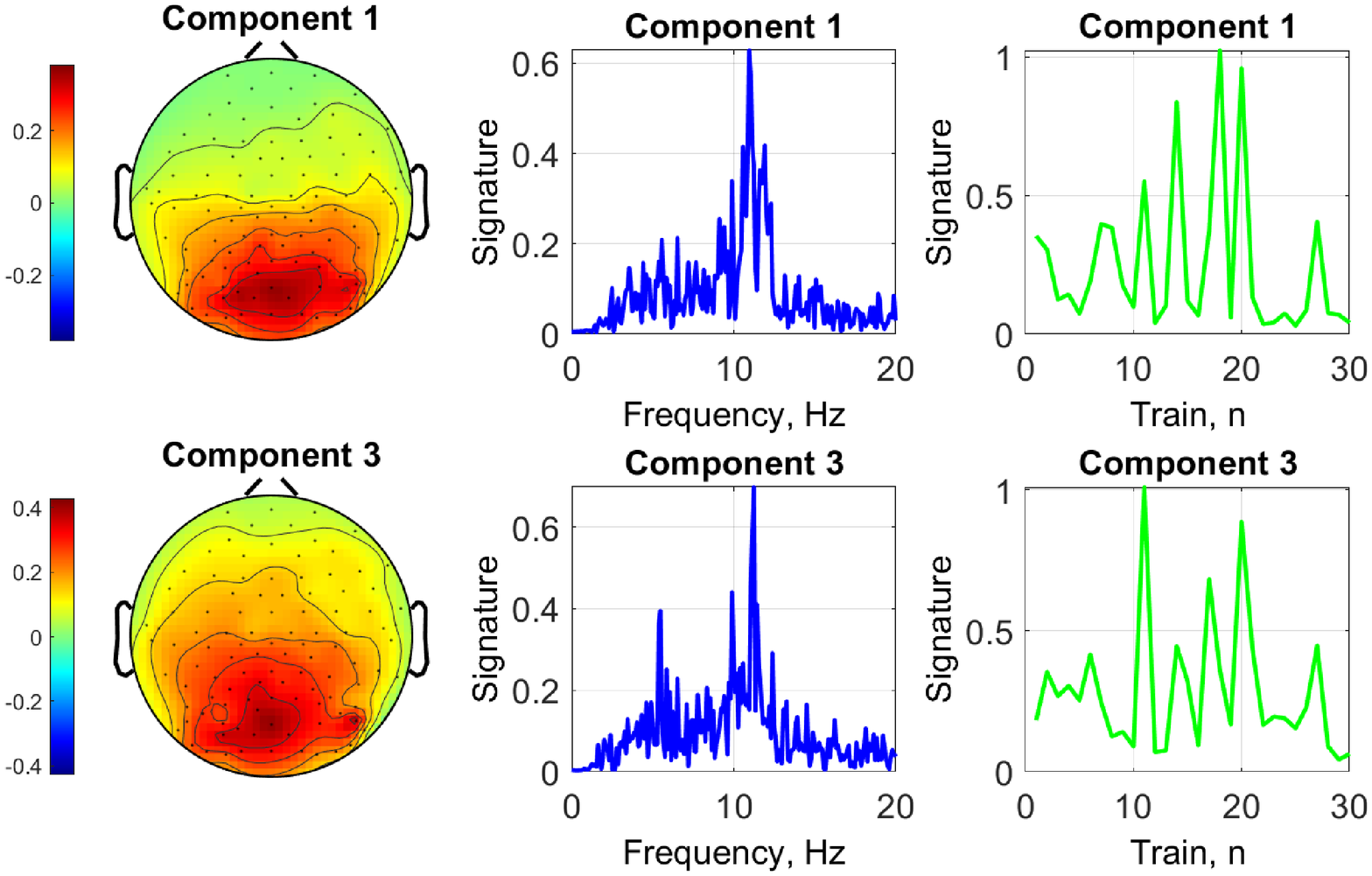} %
	\centering
	\includegraphics[width=1.08\linewidth]{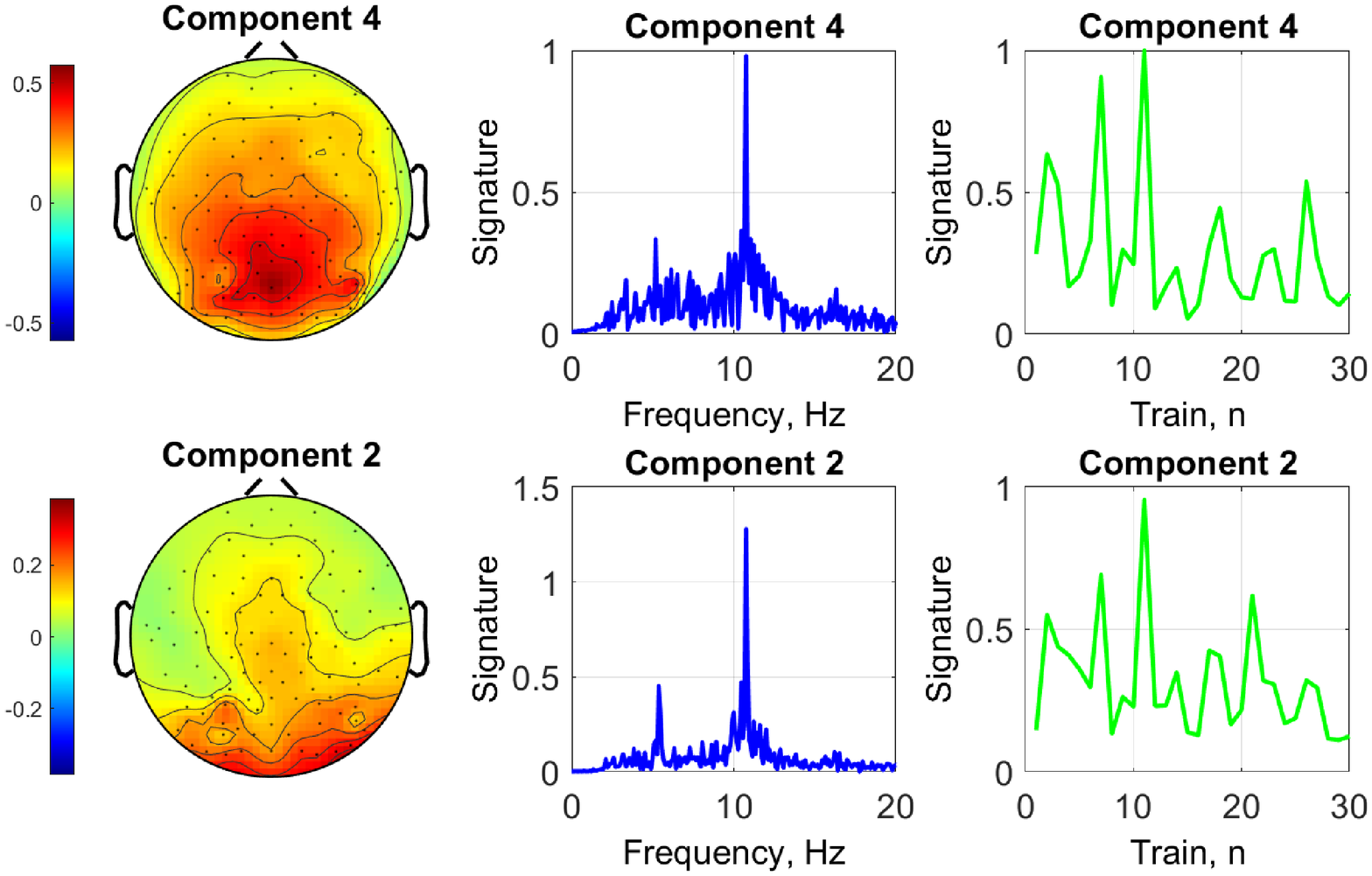} %
	\caption{Channel, frequency, and train signature for volunteer 8 and stimulation frequency 0.5$\cdot f_\alpha$ after the CP decomposition of the tensor with real EEG data $\ten{X}~\in~\compl^{106 \times 281 \times 30}$ and estimated rank 4. The loading factors of the components are $\lambda_1 = 0.2476$,$\lambda_3 = 0.2239$, $\lambda_4 = 0.2090$, $\lambda_2 = 0.1728$. }
	\vspace{-2.5mm}
	\label{fig:3D_EEG_FactMatr_Vol_8_st_freq_3_R_4}
\end{figure}

A number of interesting cases could be observed during our investigations. Fig. \ref{fig:3D_EEG_GEwPESDR_Vol_8_st_freq_3_R_4} depicts the profile of the global eigenvalues and the PESDR curves for LaRGE and LaRGE-PF. None of the PESDR coefficients computed according to the LaRGE method exceeds the threshold $\rho = 0.57$. In such a case, the model order can be estimated as one (see Figure \ref{fig:3D_EEG_FactMatr_Vol_8_st_freq_3_R_1}). Or the threshold can be decreased to the value equal to the largest PESDR coefficient. Therefore, the threshold can be updated to the new value $\rho = 0.5236$ in the considered case. Alternatively, the estimated model order based on LaRGE-PF can be used (see Figure \ref{fig:3D_EEG_FactMatr_Vol_8_st_freq_3_R_4}).


\begin{figure}[t!]
	\centering
	\includegraphics[width=1.0\linewidth]{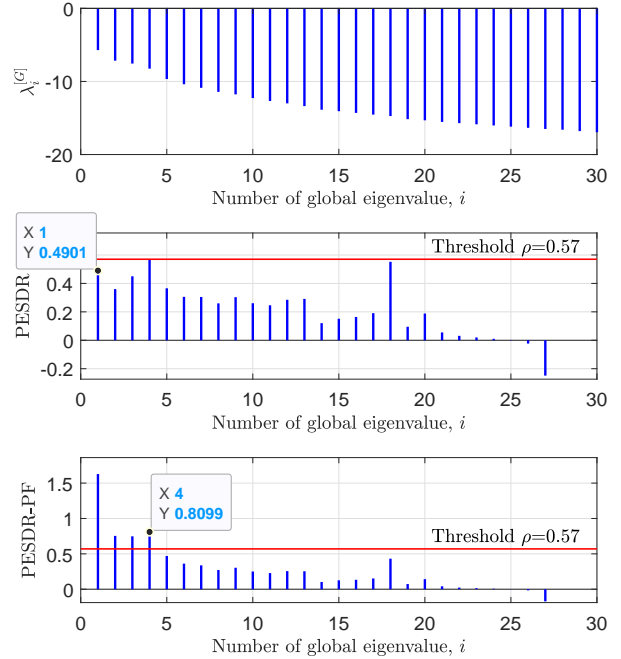} %
	\caption{The profile of the global eigenvalues (top), PESDR curve (middle) and PESDR-PF curve (bottom) computed for the tensor with real EEG data $\ten{X}~\in~\compl^{103 \times 281 \times 30}$ of volunteer 11 at stimulation frequency 0.60$\cdot f_\alpha$.}
	\vspace{-2.5mm}
	\label{fig:3D_EEG_GEwPESDR_Vol_11_st_freq_5_R_4}
\end{figure}

The profile of the global eigenvalues and the PESDR curves for a tensor with EEG data recorded from volunteer 11 at stimulation frequency $f_{ \rm{stim}} = 0.6 \cdot f_\alpha$ are depicted in Figure \ref{fig:3D_EEG_GEwPESDR_Vol_11_st_freq_5_R_4}. As in the previous case, none of the PESDR coefficients exceeds the threshold. However, the PESDR coefficients with numbers 4 and 18 are very close to the threshold. As proposed before, the model order estimated with LaRGE-PF can be used in this case.
\begin{figure}[t!]
	\centering
	\includegraphics[width=1.08\linewidth]{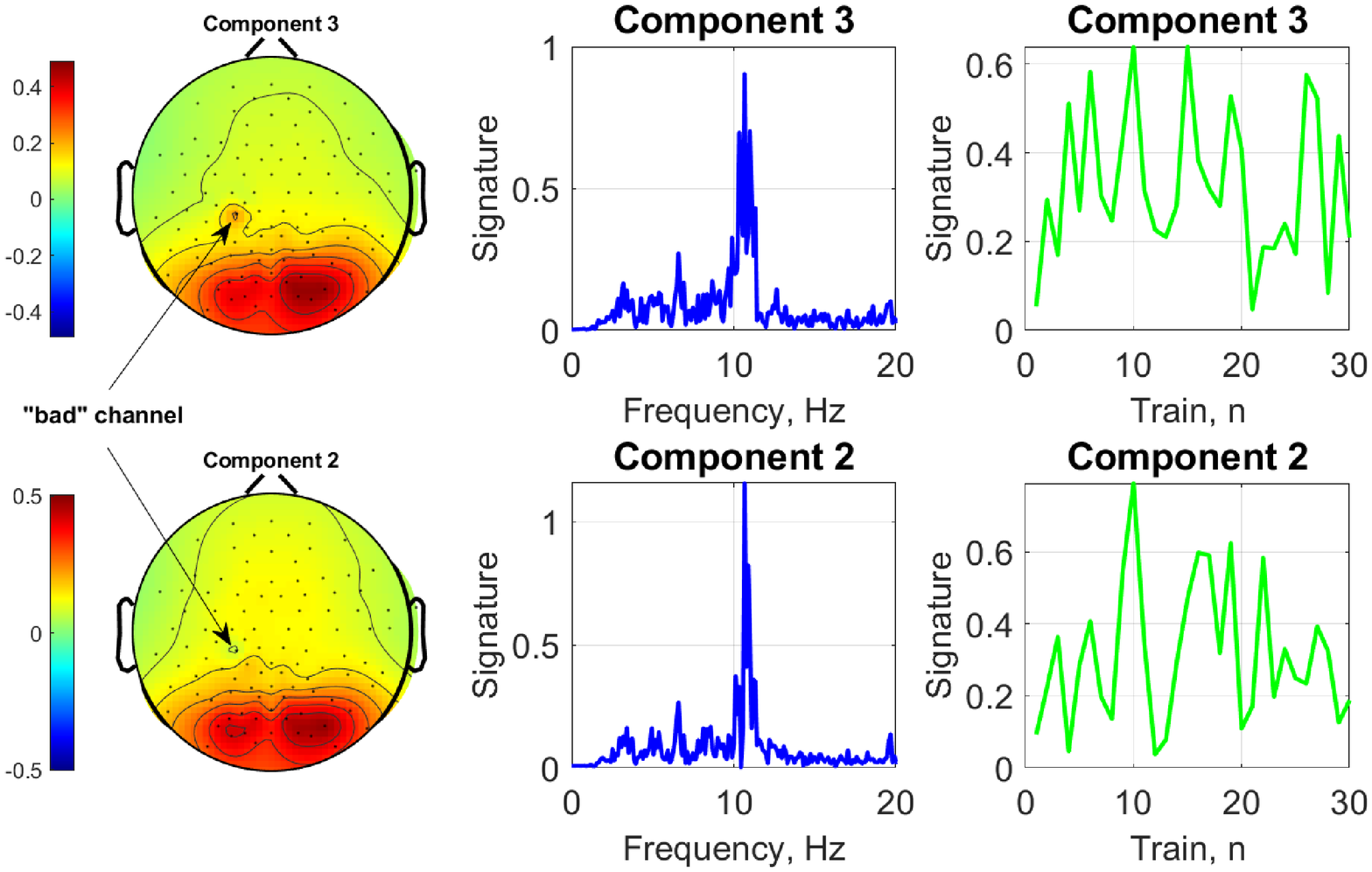} %
	\centering
	\includegraphics[width=1.08\linewidth]{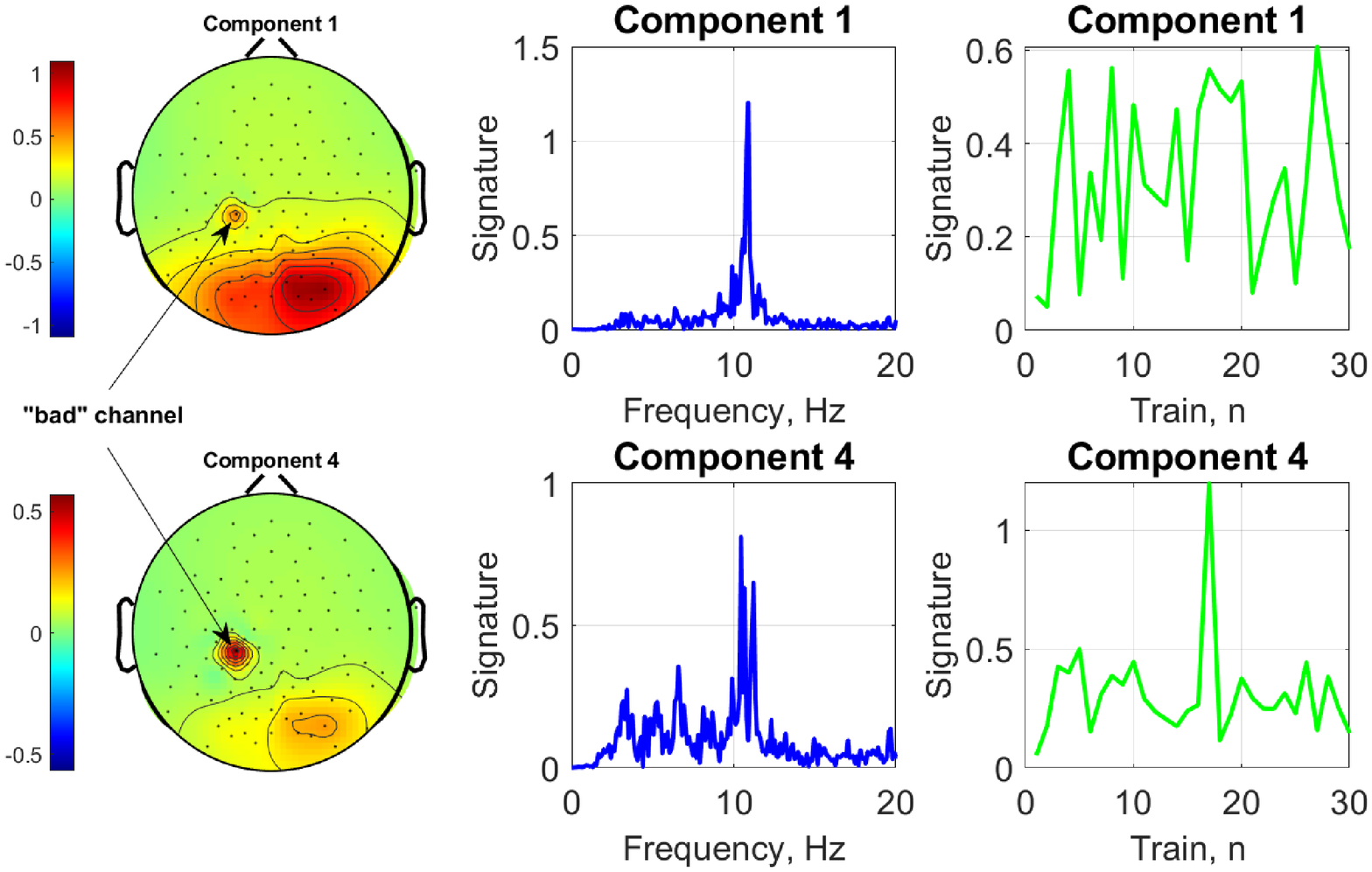} %
	\caption{Channel, frequency, and train signature for volunteer 11 and stimulation frequency 0.6$\cdot f_\alpha$ after the CP decomposition of the tensor with real EEG data $\ten{X}~\in~\compl^{103 \times 281 \times 30}$ and estimated rank 4. The loading factors of the components are $\lambda_3 = 0.2883$, $\lambda_2 = 0.2830$, $\lambda_1 = 0.2751$, $\lambda_4 = 0.2092$. The arrows indicate a "bad' channel}
	\vspace{-2.5mm}
	\label{fig:3D_EEG_FactMatr_Vol_11_st_freq_5_R_4}
\end{figure}

\begin{figure}[t!]
	\centering
	\includegraphics[width=1.0\linewidth]{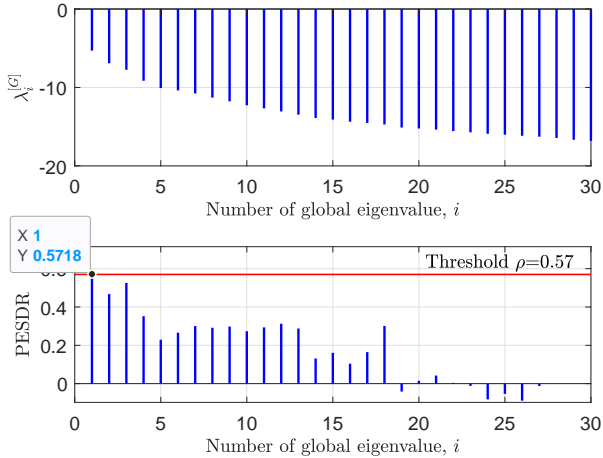} %
	\caption{The profile of the global eigenvalues (top) and PESDR curve (bottom) computed for the tensor with real EEG data $\ten{X}~\in~\compl^{102 \times 281 \times 30}$ of volunteer 11 at stimulation frequency 0.60$\cdot f_\alpha$ after "bad" channel removal.}
	\vspace{-2.5mm}
	\label{fig:3D_EEG_GEwPESDR_Vol_11_st_freq_5_R_1_afchdel}
\end{figure}

\begin{figure}[t!]
	\centering
	\includegraphics[width=1.1\linewidth]{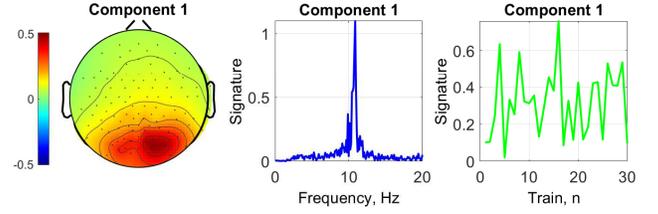} %
	\caption{Channel, frequency, time and train signature for volunteer 11 and stimulation frequency 0.60$\cdot f_\alpha$ after the CP decomposition of the tensor with real EEG data $\ten{X}~\in~\compl^{102 \times 281 \times 30}$ and estimated rank 1. }
	\vspace{-2.5mm}
	\label{fig:3D_EEG_FactMatr_Vol_11_st_freq_5_R_4_afchdel}
\end{figure}

The factor matrices obtained after the CP decomposition with rank 4 are depicted in Figure \ref{fig:3D_EEG_FactMatr_Vol_11_st_freq_5_R_4}. In all topographic plots, one "bad" channel are clearly revealed. "Bad" channel means that the corresponding sensor had an imperfect connection with the surface of the skin during the measurements. Such channels contain artifacts with low or high amplitudes and should be removed from the observations during preprocessing (which failed for this channel in this case). 
After removing this channel, the tensor was constructed and the model order was estimated again. In Figure \ref{fig:3D_EEG_GEwPESDR_Vol_11_st_freq_5_R_1_afchdel} the profile of the global eigenvalues and the PESDR curve are depicted after  removing the "bad" channel. The eighteenth PESDR coefficient becomes significantly smaller and the first coefficient exceeds the threshold now. The obtained factor matrices after channel removal and subsequent tensor decomposition are depicted in Figure \ref{fig:3D_EEG_FactMatr_Vol_11_st_freq_5_R_4_afchdel}.

\begin{figure}[t!]
	\centering
	\includegraphics[width=1.0\linewidth]{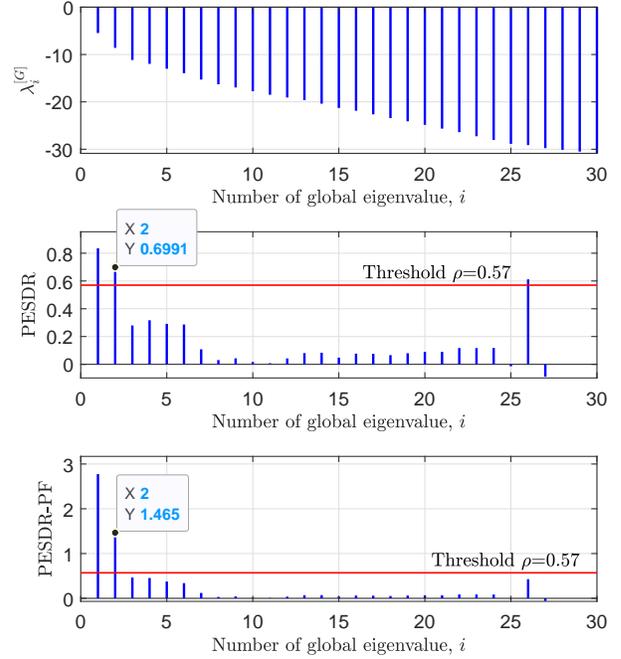} %
	\caption{The profile of the global eigenvalues (top), PESDR curve (middle) and PESDR-PF curve (bottom) computed for the tensor with real EEG data $\ten{X}~\in~\compl^{268 \times 429 \times 91 \times 30}$ of volunteer 3 at stimulation frequency 1.00$\cdot f_\alpha$ before "bad" channels removal.}
	\vspace{-2.5mm}
	\label{fig:4D_EEG_GEwPESDR_Vol_3_st_freq_10_R_2}
\end{figure}

Another extraordinary case was found for the 4-D tensors. In Figure \ref{fig:4D_EEG_GEwPESDR_Vol_3_st_freq_10_R_2}, the profile of the global eigenvalues and the PESDR curves are presented for the 4-D tensor with data that were recorded from volunteer 3 at the stimulation frequency $f_{ \rm{stim}} = 1.00 \cdot f_\alpha$. The LaRGE method overestimates the model order as 26. At the same time, LaRGE-PF estimates the model order as two. Using the topographic plots, three "bad" channels were found and removed. Then the analysis was repeated.

\begin{figure}[t!]
	\centering
	\includegraphics[width=1.0\linewidth]{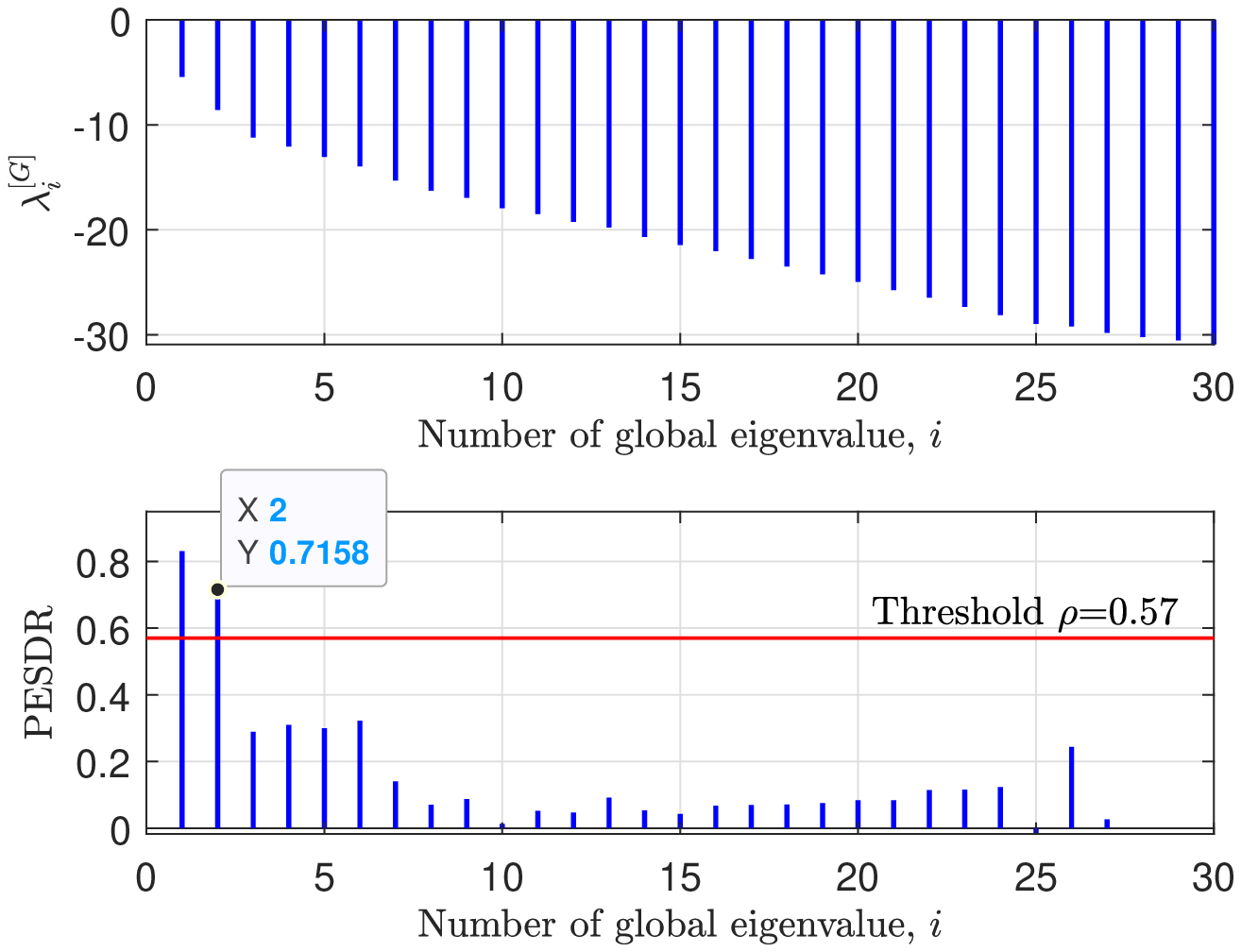} %
	\caption{The profile of the global eigenvalues (top) and PESDR curve (bottom) computed for the tensor with real EEG data Dimensions: $\ten{X}~\in~\compl^{268 \times 429 \times 88 \times 30}$ of volunteer 3 at stimulation frequency 1.00$\cdot f_\alpha$ after "bad" channels removal.}
	\vspace{-2.5mm}
	\label{fig:4D_EEG_GEwPESDR_Vol_3_st_freq_10_R_2afchdel}
\end{figure}

\begin{figure}[t!]
	\centering
	\includegraphics[width=1.1\linewidth]{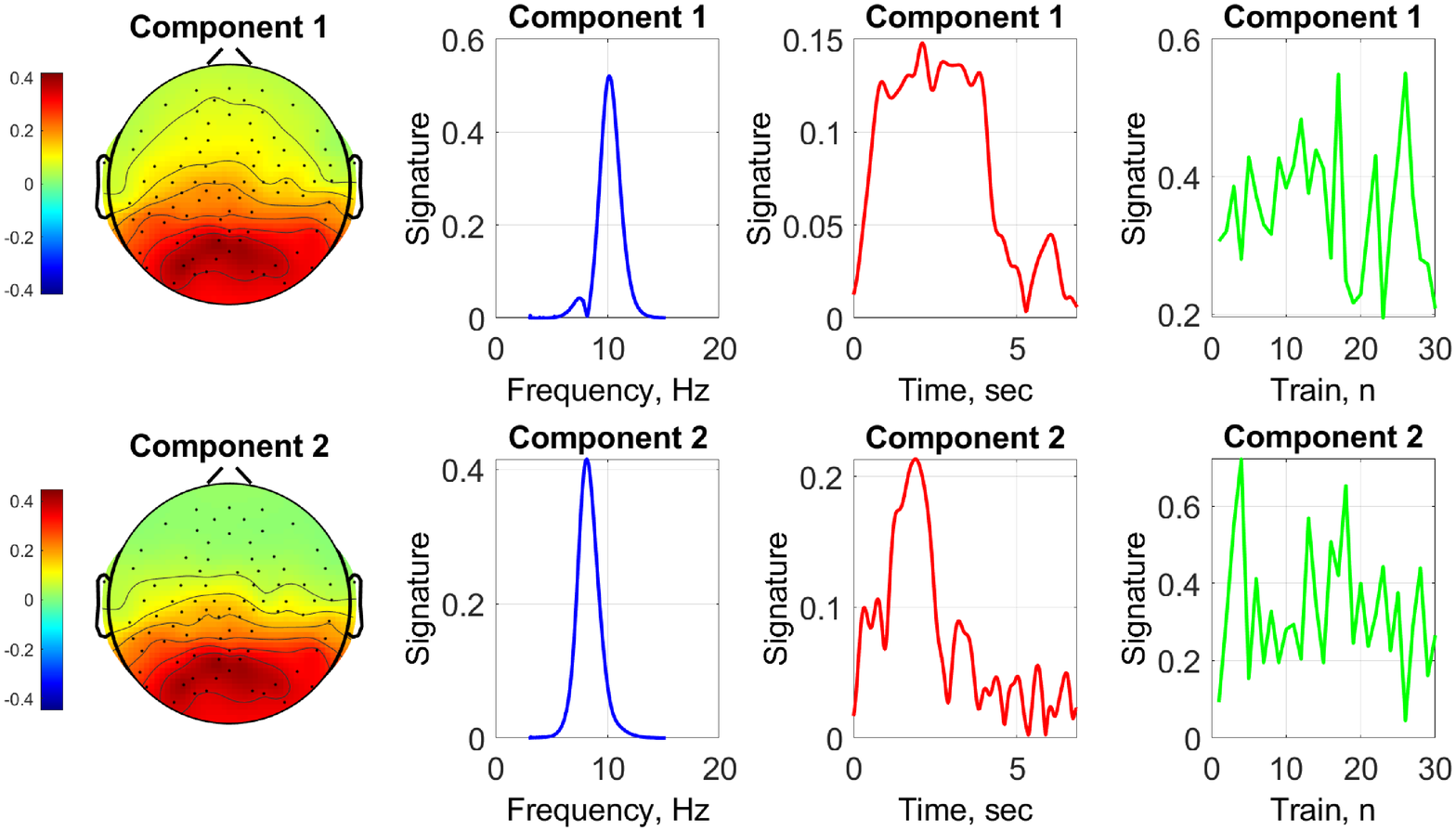} %
	\caption{Channel, frequency, time and train signature for volunteer 3 and stimulation frequency 1.00$\cdot f_\alpha$ after the CP decomposition of the tensor with real EEG data Dimensions: $\ten{X}~\in~\compl^{268 \times 429 \times 88 \times 30}$ and estimated rank 1. The loading factors of the components are $\lambda_1 = 0.3298$, and $\lambda_2 = 0.1777$ }
	\vspace{-2.5mm}
	\label{fig:4D_EEG_FactMatr_Vol_3_st_freq_10_R_2_afchdel}
\end{figure}

In Figures \ref{fig:4D_EEG_GEwPESDR_Vol_3_st_freq_10_R_2afchdel} and \ref{fig:4D_EEG_FactMatr_Vol_3_st_freq_10_R_2_afchdel} the profile of the global eigenvalues, the PESDR curve, and the resulting factor matrices after the 4-D tensor decomposition are shown. Component one represents the expected resonance effect on the individual volunteer's alpha frequency.

\section{Conclusion}
\label{sec:Conclusion}

In this paper, we have presented the robust LaRGE and LaRGE-PF methods of model order estimation based on the global eigenvalues of noise-corrupted low-rank tensors. Using the HOSVD of a measurement tensor, these global eigenvalues can be computed. Starting from the smallest global eigenvalues, the PESDR is calculated using linear regression. If it is larger than a predefined threshold $\rho$ we have found the smallest signal global eigenvalue, i.e., the approximate model order $\hat{R}$.
To prevent the misclasification in the first steps of this procedure, the modified LaRGE method with a heuristic penalty function (LaRGE-PF) has been proposed. LaRGE-PF suppresses the outliers of the PESDR curve for the tensors with small dimensions. 

Monte Carlo simulations have been conducted to determine the best value of the threshold. To this end, the probabilities of false positive, false negative, and correct detection have been defined. Moreover, we have compared LaRGE and LaRGE-PF with the specified threshold $\rho = 0.57$ with classical AIC, MDL, 3-D AIC, 3-D MDL, 4-D AIC, and 4-D MDL methods. LaRGE and LaRGE-PF show an increased robustness and outperform the classical methods significantly. 

In this paper, we have also used the LaRGE and LaRGE-PF methods to estimate the model order of measured EEG data during an experimental investigation of the photic driving effect. The dominant components can be reliably detected by exploiting the PESDR curve. 

If "bad" channels have been identified from the extracted dominant components, PESDR coefficients with a high amplitude could be observed. This effect might be very important for practical applications such as the automatic detection and removal of artifactual EEG channels and could be investigated in future research.
Moreover, LaRGE could be extended to coupled tensors decompositions. 
\ifCLASSOPTIONcaptionsoff
  \newpage
\fi



%
\bibliographystyle{IEEEtran}

\bibliography{mybibfile}
\end{document}